\documentclass[12pt]{article}
\usepackage{amsmath}
\usepackage{graphicx,epsfig}
\usepackage{amssymb}
\usepackage{subfigure}
\usepackage[margin=1in]{geometry}
\usepackage{tabu}
\usepackage{cite}
\usepackage{caption}
\usepackage{url}
\def\beq {\begin{equation}}
\def\eeq {\end{equation}}
\def\bea {\begin{eqnarray}}
\def\eea {\end{eqnarray}}
\def \PMET{\rm p{\!\!\!/}_T}

\def\noi{\noindent}
\def\G{\Gamma}

\newcommand{\br}{\begin{eqnarray}}
\newcommand{\er}{\end{eqnarray}}
\newcommand{\be}{\begin{equation}}
\newcommand{\ee}{\end{equation}}

\newcommand{\stopo}{\widetilde t_1}
\newcommand{\sboto}{\widetilde b_1}

\newcommand{\ra}{\rightarrow}
\newcommand{\mgino}{m_{\tilde g}}
\newcommand{\gino}{\tilde g}
\newcommand{\wino}{\widetilde{W}}
\newcommand{\bino}{\widetilde{B}}
\newcommand{\higgino}{\widetilde{H}}
\newcommand{\mbino}{m_{\bino}}

\newcommand{\mhiggino}{m_{\higgino}}
\newcommand{\neut}[1]{{\tilde \chi}_{#1}^0}
\newcommand{\charg}[1]{{\tilde \chi}_{#1}^\pm}
\newcommand{\mneut}[1]{m_{{\tilde \chi}_{#1}^0}}
\newcommand{\mcharg}[1]{m_{{\tilde \chi}_{#1}^\pm}}

\usepackage{color}
\usepackage{hyperref} 
\hypersetup{linktocpage=true}
\usepackage[all]{hypcap}
\hypersetup{
   bookmarks=true,         
   unicode=false,          
   colorlinks=true,       
   linkcolor=blue,          
   citecolor=magenta,        
   filecolor=magenta,      
   urlcolor=cyan           
}
\usepackage{fancyhdr}
\begin{document}
\thispagestyle{empty}
\vspace*{-22mm}
\vspace*{10mm}
\hypersetup{backref=true,bookmarks}

\vspace*{10mm}
\begin{flushright}
 LPSC-15-224
\end{flushright}

\begin{center}
{\Large {\bf\boldmath 
Closing in on compressed gluino-neutralino spectra at the LHC
}}

\vspace*{10mm}

{\bf Guillaume Chalons, Dipan Sengupta}\\
\vspace{4mm}
{\small

Laboratoire Physique Subatomique et Cosmologie, Universit\'{e} Grenoble-Alpes,\\
 CNRS/IN2P3, Grenoble INP, 53 rue des Martyrs, F-38026 Grenoble, France
}

\vspace*{10mm}

{\bf Abstract}\vspace*{-1.5mm}\\
\end{center}
A huge swath of parameter space in the context of the Minimal Supersymmetric Standard Model (MSSM) has been ruled at after run I of the LHC. Various exclusion contours in the 
$\mgino-\mneut{1}$ plane were derived by the experimental collaborations, all based on 
three-body gluino decay topologies. These limits are however extremely model dependent and 
do not always reflect the level of exclusion. If the gluino-neutralino spectrum is 
compressed, then the current mass limits can be drastically reduced. In such situations, 
the radiative decay of the gluino $\gino \ra g \neut{1}$ can be dominant and used as a 
sensitive probe of small mass splittings. We examine the sensitivity of constraints of 
some Run I experimental  searches on this decay after recasting them within the 
\texttt{MadAnalysis5} framework. The recasted searches are now part of the 
\texttt{MadAnalysis5} Public Analysis Database. We also design a dedicated search strategy 
and investigate its prospects to uncover this decay mode of the gluino at run II of the 
LHC. We emphasize that a multijet search strategy may be more sensitive than a monojet 
one, even in the case of very small mass differences.

\vspace*{10mm}


\noindent

\newpage

\tableofcontents


\section{Introduction and motivation}
During run I of the LHC, in addition to the quest for the Higgs boson,
 an extensive search programme for New Physics (NP) phenomena has been carried out. 
However, no sign of physics beyond the Standard Model (BSM) has been unravelled so far 
\cite{atlas-susy-twiki,cms-susy-twiki,atlas-exo-twiki,cms-exo-twiki}.
As a result the parameter space for weak scale NP, including models
 of weak scale Supersymmetry (SUSY), has shrunk considerably. 
\par\noi 
The simplest incarnations of weak scale SUSY, the MSSM and its variants,
 depending on the SUSY breaking scheme, are facing a wealth of experimental 
data~\cite{atlas-susy-twiki,cms-susy-twiki,atlas-exo-twiki,cms-exo-twiki}, 
including the Higgs 
boson mass measurement $m_H \simeq 125$ GeV \cite{Aad:2015zhl}. Since no NP signal was revealed,
 the only appropriate physical interpretation of these searches was to set
 limits on cross sections for the production of SUSY particles and then derive
 bounds on their masses. Interpreting the NP experimental searches is a non-trivial
 task and almost impossible to perform in a model-independent way.
 The early searches for SUSY phenomena have focussed on the constrained MSSM (CMSSM)/minimal
 supergravity (mSUGRA) and probed 
mainly the gluino ($\gino$) and the squarks ($\tilde q$) of the first two generations 
\cite{Aad:2015iea,deJong:2012zt}. 
The current lower limits on the gluino and the first two generations squark masses in the 
framework of the CMSSM stand at $\mgino \gtrsim 1.7$ TeV for almost degenerate gluino and 
squarks and $m_{\tilde q} > 1.4$ TeV for very high squark masses. However, the universal 
boundary conditions imposed at the Grand Unified Theory (GUT) scale in such models impose 
particular relations between particle masses, decay branching ratios, etc... which severely 
restrict the way a SUSY signal could show up. To relax some of these assumptions, 
the Simplified Model Spectra (SMS) 
\cite{Okawa:2011xg,Chatrchyan:2013sza,Alwall:2008ag,Alves:2011sq,Alves:2011wf,Ghosh:2012wb} approach 
has been adopted systematically by the ATLAS and CMS collaborations to interpret the 
results. Instead of considering the full SUSY spectrum, a SMS search only targets one or 
two relevant decay modes (and the particles involved in) with 100\% branching ratio (BR). 
In turn, a handful of parameters are to be considered for interpreting the analysis: the 
production cross section of the mother particle, its mass, and the ones of the daughters. 
Nevertheless, such SMS approaches may not cover all patterns in which a NP signal could
be unveiled, especially if one consider the plethora of existing supersymmetric models.
 More precisely, even in R-parity conserving scenarios, if the lightest SUSY particle (LSP)
 is massive and degenerate with the squarks and/or gluinos, the so-called ``compressed'' SUSY scenarios,
 these limits can be seriously weakened \cite{Dreiner:2012gx,Alwall:2008ve,Izaguirre:2010nj,Alwall:2008va,LeCompte:2011cn,LeCompte:2011fh,Bhattacherjee:2012mz}. 
\par\noi 
As far as the official experimental analyses are concerned, published 
gluinos mass limits in the SMS approach were obtained mainly from its three-body decays 
$\gino \ra q \bar{q} \neut{1}$, where the final states quark can be of any flavour 
\cite{Aad:2015iea,Aad:2014lra,Aad:2013wta,Chatrchyan:2013iqa,Chatrchyan:2013fea,
CMS:2014wsa} or $\gino \ra q q'  \charg{1}$ \cite{Aad:2015iea,Aad:2014lra}. Provided the 
gluino-neutralino spectrum is compressed, the limits derived from three-body decays 
involving heavy-flavours do not apply and only the ones with light-flavours in the final 
states can give a handle on this situation. However, as limits on the first and second generation of squarks are quite strong, the $\gino \ra q \bar{q} \neut{1}$ modes with light quarks may be heavily suppressed. 
Furthermore, as already stressed,  a large part of the parameter space feature a 
compressed 
spectra to evade the stringent bounds from the LHC, and few SMS 
interpretations fully encompass this situation for the gluino. In compressed scenarios, 
initial state radiation (ISR) in the form of jets which recoil against the missing 
momentum of the LSP may be used to discover or place bounds on the model. This method has 
been used to set mass limits on squarks and gluinos 
\cite{Alwall:2008ve,Alwall:2008va,Izaguirre:2010nj,LeCompte:2011cn,LeCompte:2011fh,
Dreiner:2012gx,Bhattacherjee:2012mz,Carena:2008mj,Alvarez:2012wf,Dreiner:2012sh,
Arbey:2015hca,Aad:2015iea,Chatterjee:2014uda} and third generation squarks 
\cite{Dreiner:2012gx,He:2011tp,Drees:2012dd,Arbey:2015hca,Aad:2015pfx,CMS:2014yma,
Ferretti:2015ala}.
\par\noi 
In the case where the three body decays are kinematically forbidden or suppressed, there exists 
another decay pattern available for the gluinos to decay into, which has received very 
little attention so far, but could fill the small mass difference gap: that of the two-body radiative decay 
$\gino \ra g \neut{i}$.
\par\noi 
In a CMSSM-inspired scenario, since the mass difference $\Delta M = \mgino - \mneut{1}$ is 
fixed from conditions imposed at the GUT scale, when $\mgino$ increases, so does $\Delta 
M$. In turn, the decay width of the gluino will be dominated by the three-body decays. Unlike GUT 
constrained SUSY-scenarios, general explorations of the MSSM may exhibit significant 
portions of parameter space where the radiative decay $\gino \ra g \neut{1}$ can be 
competitive with three-body decays 
\cite{Toharia:2005gm,Ma:1988ns,Barbieri:1987ed,Haber:1983fc,Baer:1990sc,Bartl:1990ay,
Gambino:2005eh,Sato:2012xf,Ghosh:2012dh,Chatterjee:2012qt}. Within these specific regions, significant branching 
fractions of the gluino radiative decay are characterised by a spectrum with heavy 
squarks. Thus, the direct two-body decays $\gino \ra \tilde{q} q$ are forbidden and $\gino 
\ra q \bar{q} \neut{i}$ are suppressed by the heavy scalar quark mass, leaving the 
loop-induced decay as the only available mode to decay into. Therefore, for 
compressed gluino-neutralino scenarios, the gluino loop decay can be used as a sensitive 
probe and improve the current limits.
\par\noi 
A spectrum with decoupled scalars and lighter gauginos and higgsinos is not unexpected in 
SUSY models like (mini) Split-SUSY/PeV scale SUSY 
\cite{Wells:2003tf,Wells:2004di,ArkaniHamed:2004yi,Giudice:2004tc,ArkaniHamed:2004fb,
Arvanitaki:2012ps,Harigaya:2013asa},
 Spread SUSY scenarios \cite{Hall:2011jd} or pure gravity mediation \cite{Ibe:2011aa,Ibe:2012hu,Harigaya:2013asa,Evans:2014xpa}. Moreover, a Higgs mass
 around 125 GeV implies a non-negligible branching fraction of the gluino loop decay
 when $\tan \beta \simeq {\cal O}(1)$ \cite{Sato:2012xf}. It is worth mentioning that if
 the scalar superpartners are heavy enough, then the gluino lifetime $\tau_{\tilde g}$ can
 be sufficiently large to lead to displaced vertices or even hadronise into R-hadrons \cite{ArkaniHamed:2004fb,Giudice:2004tc,Aad:2013gva,ATLAS:2014fka}. Nevertheless,
 if $\tilde m \lesssim {\cal O}(10^4)$ TeV, $\tau_{\tilde g}$ is too small such that the gluino
 decays promptly in the detector \cite{Sato:2012xf}. 
\par\noi 
 To obtain a compressed gluino-neutralino spectrum, given a SUSY-breaking scheme, one needs non-universalities in the gaugino sector at the GUT scale to obtain a small $\Delta M$. In the context of non-universal supergravity (NUSUGRA) models, the gluino can be in a significant part of the parameter space the next-to-lightest SUSY particle (GNLSP) and the neutralino the LSP \cite{Feldman:2007fq,Feldman:2008hs,Feldman:2009zc,Chen:2010kq,Baer:2006dz,Baer:2006ff,Baer:2008ih,Guchait:2011hj}.
 A non-exhaustive list of other SUSY models accommodating the GNLSP scenario include 
left-right symmetric model with gravity mediated SUSY 
breaking \cite{Gogoladze:2009ug,Gogoladze:2009bn,Ajaib:2010ne,Raza:2014upa}, pure gravity mediation/mini-split SUSY \cite{Harigaya:2013asa,Evans:2014xpa} and general gauge mediation \cite{Grajek:2013ola}.
\par\noi 
Besides being a discovery mode for the gluino, its loop-decay has some other merits from 
the phenomenological perspective. Indeed, since squarks contribute to the radiative decay 
through loop diagrams, the two-body decays $\gino \ra g \neut{i}$ carry important 
information about the scalar mass scale $\tilde m$ \cite{Sato:2012xf}, which would be 
otherwise inaccessible in scenarios with decoupled scalars. Moreover, in GNLSP models, the 
LSP can account for the relic density of dark matter through its coannihilation with the 
gluino. The efficiency of such processes crucially depends on the gluino-LSP mass 
difference $\Delta M$. Thus, measuring or constraining this splitting, only accessible at 
collider experiments, is essential to estimate the phenomenological viability of these 
models.
\par\noi 
We should also note in passing, that in models with spontaneously broken global supersymmetry the radiative decay
$\gino \ra g \widetilde G$, where $\widetilde G$ is the gravitino, is also possible. It 
can be dominant when the original scale of supersymmetry breaking $\sqrt{F}$ is of the 
same order of the squark mass scale $\widetilde m$ 
\cite{Dicus:1989gg,Drees:1990vj,Dicus:1996ua,Kim:1997iwa,Brignole:1998me,
ArkaniHamed:2004yi,Babu:2005ui,Gambino:2005eh,Klasen:2006kb,deAquino:2012ru}, whatever the 
$\mgino -m_{\widetilde G}$ mass difference. Depending on the value of $F$ the gluino can 
decay promptly, in flight inside the detector, or be long-lived. For very long-lived 
gluinos, ATLAS and CMS have excluded charged R-hadrons with a mass less than 1.3 TeV 
\cite{Aad:2012pra,Chatrchyan:2013oca}, if they escape the detector, or less than 850 GeV 
\cite{Aad:2013gva} if stopped in the detector. If metastable, the gluino is excluded up 
to $\mgino = 850$ GeV for decays to $q \bar{q} \neut{1}/g\neut{1}$, for a lifetime of 1 
ns and $\mneut{1} = 100$ GeV \cite{ATLAS:2014qga}. 
\par\noi
A work probing scenarios where the LSP and the gluino are almost degenerate was recently performed in \cite{Arbey:2015hca}. However, the limits derived there solely considered the three-body decays of the gluino into 
light-flavour quarks $\gino \ra q \bar{q} \neut{1}$, but with light sbottoms and light 
stops in the spectrum. As we will review in the forthcoming sections, the radiative 
decay $\gino \ra g \neut{1}$ is favoured when light stops are present and should dominate 
instead. Therefore, investigating rather the gluino loop-induced decay seems better motivated from a 
phenomenological point of view when all scalars are heavy and $\Delta M$ small. The ATLAS 
collaboration has also recently provided, using 
only a hadronic search that we reimplemented for this work, some results on the radiative 
gluino decay in \cite{Aad:2015iea}. Yet, results have only been presented for a fixed 
gluino mass $\mgino$ or fixed LSP mass $\mneut{1}$, but not in the full 
$\mgino-\mneut{1}$ 
plane. 
\par\noi 
In this work, we aim at quantifying to which extent the radiative gluino decay can 
constrain, using Run I monojet and multijets analyses, the region where the mass splitting 
$\Delta M = \mgino - \mneut{1}$ is small, assuming a SMS interpretation where 
$\mbox{BR}(\gino \ra g \neut{1}) = 100\%$. We argue in the next section, after 
investigating the parameter space where the loop decay is important, why a somewhat 
lighter third generation squark spectrum is desirable to enhance this decay from the 
phenomenological MSSM (pMSSM) point of view.  We use the \texttt{MadAnalysis5} 
\cite{Conte:2012fm,Conte:2014zja} framework to derive exclusion contours in the 
$\mgino-\mneut{1}$ plane, deduced from a Simplified Model approach, after recasting ATLAS 
and CMS monojet and multijets analyses. We provide in our work a full 
quantitative exploration of this plane and the quoted 
ATLAS 95\% CL upper bound \cite{Aad:2015iea} for fixed $\mgino$ and $\mneut{1}$ rather 
serves to us as a 
cross check on our results, which give us quite confidence in our procedure with which we 
derived our own mass limits. These recasted analyses are now available on the 
\texttt{MadAnalysis5} Public Analysis Database (PAD) \cite{Dumont:2014tja}. The last 
section is devoted to the determination of a search strategy at 13 TeV to investigate the 
prospects for discovering the gluino through its radiative decay at run II of the LHC and 
then we conclude.

\section{Phenomenology of the radiative gluino decay}
\label{pheno}
\begin{figure}[h]
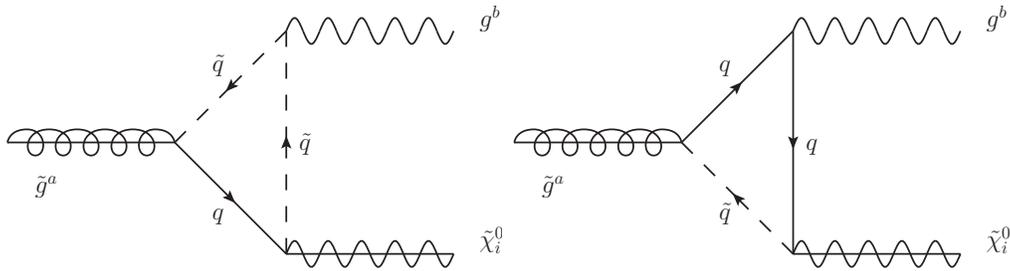

 \begin{center}
  \epsfig{file=Diag1,clip=,width=.4\textwidth}
  \epsfig{file=Diag2,clip=,width=.4\textwidth}
 \end{center}\caption{\em \label{fig:diagraddec}Diagrams relevant for the
radiative gluino decay. Since the gluino is a Majorana fermion each diagram is
accompanied with a similar one but with the internal arrows reversed.}
\end{figure}

In this section we review the gluino decays in a heavy scalar scenario. When the
squarks are heavier than the gluino, the latter can undergo three body decays
into a pair of quarks and a neutralino or chargino through an off-shell intermediate squark, or decay radiatively
into a gluon and a neutralino,
\begin{align}
 \gino &\ra q + \bar q + \neut{i} \\
 \gino &\ra q + \bar q' + \charg{j} \\
 \gino &\ra g + \neut{i}
\end{align}\noi
where $\neut{i}$ ($i=1\dots4$) and $\charg{j}$ ($j=1,2$) are neutralinos and charginos 
respectively. Let
us first review the radiative two body decays of the gluino, which occur via
quark-squark loop diagrams (Fig.~\ref{fig:diagraddec}).
\noi 
The neutralino $\neut{i}$ is an admixture of wino- ($\wino_3$), bino-
($\widetilde B$) and higgsino-like ($\widetilde H$) neutral spinors, which are 
respectively the superpartners of the neutral gauge eigenstates $B,W_3$  and of the two 
Higgs doublet of the MSSM $H_{u,d}$,
\begin{equation}
 \neut{i} = N_{i1} \bino + N_{i2} \wino_3 + N_{i3} \higgino_d + N_{i4}
\higgino_u
\end{equation}\noi 
If the squarks are very heavy, from an effective Lagrangian point of view the
decay width into a wino is strongly suppressed since it is induced by a
dimension 7 operator \cite{Gambino:2005eh,Sato:2012xf}, whereas for a bino or a
higgsino it can be induced already by a dimension 5 chromo-magnetic interaction,
\begin{equation}
\label{dim5gluino}
 {\cal L}_{\rm eff.} = \frac{1}{\widetilde m}\overline{\neut{i}} \sigma^{\mu\nu}
P_{L,R} \gino^a G_{\mu\nu}^b \delta_{ab}
\end{equation}\noi 
where $\widetilde m$ is an effective squark mass scale. The partial widths of
these two body-decays are given by \cite{Toharia:2005gm,Gambino:2005eh,Ajaib:2010ne,Sato:2012xf}, in the heavy squark limit,
\begin{eqnarray}
\label{twobodybino}
 \G(\gino \ra g \bino) &\simeq&
\frac{\alpha \alpha_s^2}{512 \pi^2 c_W^2}\frac{\left(\mgino^2-\mbino^2\right)^3}{\mgino^3
}\left[\sum_q \frac{Y_{q_L}}{m^2_{{\widetilde
q}_L^2}}-\frac{Y_{q_R}}{m^2_{{\widetilde
q}_R^2}}\right]^2\left(\mgino-m_{\widetilde B}\right)^2
 \\
\label{twobodyhiggino}
 \G(\gino \ra g \higgino)
&\simeq&\frac{\alpha \alpha_s^2 m_t^2 }{128 \pi^2 M_W^2 s_W^2 s_\beta^2}
\frac{\left(\mgino^2-\mhiggino^2\right)^3}{\mgino^3}
\left[\frac{m_t}{m^2_{\tilde
t_L}}\left(\ln\frac{m^2_{\tilde t_L}}{m_t^2}-1\right)+\frac{m_t}{m^2_{\tilde
t_R}}\left(\ln\frac{m^2_{\tilde t_R}}{m_t}-1\right)\right]^2
\end{eqnarray}\noi
where $\alpha$ and $\alpha_s$ are the electromagnetic and strong couplings constants 
respectively. We used the short-hand notations $c_W \equiv \cos \theta_W, s_W \equiv 
\sin\theta_W, s_\beta \equiv \sin \beta$. The angle $\theta_W$ is the weak mixing angle 
and $\beta$ is defined through the ratio $\tan \beta = v_u/v_d$ where $v_{u,d}$ are the 
vacuum expectation values of the two $SU(2)_L$ Higgs doublets $H_{u,d}$ of the MSSM. Both 
angles lie between $0$ and $\pi/2$. $Y_q$, appearing in Eq.~\ref{twobodybino}, is the 
quarks hypercharge  where the index $q$ runs through all flavours. In 
Eq.~\ref{twobodyhiggino}, since the decay width is proportional to the top mass, only the 
stops-tops loops were kept. From Eq.~\ref{twobodybino} we can see that $\G(\gino \ra g 
\bino)$ scales like
$m_{\tilde q}^{-4}$ whereas $\G(\gino \ra g \higgino)$ benefits from a
logarithmic enhancement. The origin of this logarithmic enhancement can be 
understood from an effective Lagrangian point of view as follows. In the effective theory
where the squarks have been integrated out, this decay is induced by a
top-top-gluino-higgsino operator in which the two top quarks form a loop which
can emit a gluon to form the chromo-magnetic gluino-gluon-higgsino interaction of Eq.\ref{dim5gluino}. Such a
diagram is divergent in the effective theory thereby generating a logarithmic
enhancement \cite{Toharia:2005gm,Gambino:2005eh}, which is cut-off by the
squark mass (the scale of the effective theory breakdown). Therefore this decay
is an interesting probe of the scalar mass scale in case they are
kinematically inaccessible from colliders \cite{Sato:2012xf}. In the very heavy squark limit the leading
logarithmic corrections should be resummed to obtain reliable predictions using renormalisation-group techniques
as has been done in \cite{Gambino:2005eh}. 
\par
\begin{figure}[h]
 \begin{center}
  \epsfig{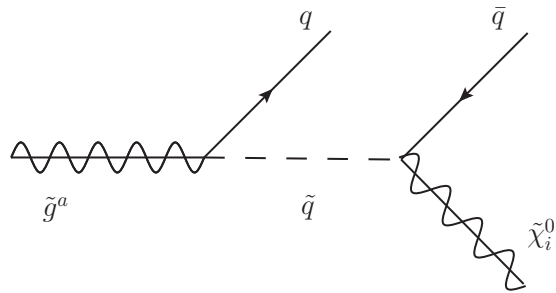}
 \end{center}\caption{\em \label{Diag3bod}Three-body decay of the gluino where the (s)quark flavour runs through $q=u,d,c,s,t,b$.}
\end{figure}

\noi 
Let us now turn on to the three body decays of the gluino $\gino \ra q \bar q
\neut{}$ which is mediated by squark exchange $\tilde q$ (Fig.\ref{Diag3bod}), with the assumption that the squark sector respects flavour symmetries.
\noi 
 In the massless quark limit, only the bino-like contributes and we can get an analytical expression of the
gluino partial decay width \cite{Haber:1983fc,Gambino:2005eh,Ajaib:2010ne,Sato:2012xf},
\begin{equation}
 \G(\gino \ra q_L q_R^c \bino) = \frac{\alpha \alpha_s  Y_{q_L}}{96 \pi \cos^2 \theta_W}
\frac{m_{\gino}^5}{m^4_{\tilde
q_L}}\left[f\left(\frac{m^2_{\bino}}{m^2_{\gino}}\right)+ \frac{2
m_{\bino}}{m_{\gino}}g\left(\frac{m^2_{\bino}}{m^2_{\gino}}\right) \right]
\end{equation}\noi 
where $f(x) = 1 - 8x -12x^2 \ln x + 8 x^3-x^4$ and $g(x) = 1 + 9x + 6x \ln x -9
x^2+ 6x^2 \ln x -x^3$. Like the two-body decay into the bino-like neutralino 
in Eq.~(\ref{twobodybino}), the three-body decay width scales as $m_{\tilde q_L}^{-4}$ in
the heavy squark limit. If the first two squarks generations are decoupled from
the third one, which is quite easily achieved in SUSY since the large top Yukawa
usually drives the running of third generation squark masses to lighter value than
the two others, the quarks produced in the three body decays will be composed of
$t\bar t, b \bar b$ pairs. For generic quark masses the integration
over the three-body phase space cannot be performed analytically and one must resort to a numerical integration. However, keeping only the quark mass
dependence in the couplings and in the limits $m_{\neut{}} \ll m_{\gino}$
and $m_{\gino} \ll m_{\tilde q_L},m_{\tilde q_R}$, the decay rate $\gino \ra q
\bar q \neut{i}$\cite{Ma:1988ns} is given by
\begin{equation}
\label{lim3bodMa}
 \G(\gino \ra q \bar q \neut{i}) \simeq \frac{\alpha_s m_{\gino}^5}{384
\pi^2}\sum_q \left[\frac{|a_q|^2 + |b_{q_L}|^2}{m_{\tilde q_L}^4}+
\frac{|a_q|^2+|b_{q_R}|^2}{m_{\tilde q_R}^4} \right]
\end{equation}\noi 
The sum is over all quark species allowed by kinematics. The $a,b,c$
coefficients are given by
\begin{align}
 a_u &= \frac{g m_u N_{i4}}{2 M_W \sin \beta},\quad a_d = \frac{g m_d N_{i3}}{2 M_W
\cos \beta},\\
\quad b_{q_L} &= g\left( \frac{Y_{q_L}}{2}\tan \theta_W N_{i1} +
T_3^{q_L} N_{i2}\right), \quad b_{q_R} = g Q_q \tan \theta_W N_{i1}
\end{align}\noi 
where $u \equiv u,c,t$ and $d \equiv d,s,b$. The electric charge $Q_q$ is defined as $Q_q = 
T_3^{q_L} + Y_q/2$.  It is then natural to ask oneself in which cases the two-body decay 
can
dominate over the three-body one. To investigate this we define the ratio
$R_{2/3}$ as
\begin{equation}
 R_{2/3} = \frac{\G(\gino \ra g \neut{i})}{\G(\gino \ra q \bar q \neut{i})} 
\end{equation}\noi
Obviously, the three-body decay is suppressed by the phase space when the mass
difference between the gluino and the neutralino is too small.  This is
particularly true for $\gino \ra t \bar t \neut{i}$ when the $t \bar t$
threshold is closed, {\it i.e} $m_{\gino}-m_{\neut{i}} < 2 m_t$. In this
situation the decay rates for $\gino \ra b \bar b \neut{i}$ may still compete with 
$\G(\gino \ra g \neut{i})$. In the higgisno-like case, the $b \bar b$ and $g \neut{i}$ 
final states are dominated by the heaviest quark: the former is proportional to $m_{b}^2$ 
and the latter to $m_t^4$. In the limit where the left and right squark masses are equal 
for each respective flavour $q$, \textit{i.e} $m_{\tilde q_L}=m_{\tilde q_R}= \tilde 
m_q$, using Eqs.~(\ref{twobodyhiggino}) and (\ref{lim3bodMa}) and defining $x_t = 
m_t^2/{\tilde m}_t^2$, one finds for $R_{2/3}$ \cite{Ma:1988ns}
\begin{equation}
\label{Rtwothree}
 R_{2/3} = \frac{24 \alpha_s}{\pi}\left(\frac{{\tilde m}_b}{{\tilde
m}_t}\right)^4\left(\frac{m_t^2}{\mgino m_b \tan
\beta}\right)^2\left[\frac{1}{1-x_t}+ \frac{\ln x_t}{(1-x_t)^2}\right]^2
\end{equation}\noindent
In the bino-like case, we would obtain a similar formula using Eqs.~(\ref{twobodybino}) and (\ref{lim3bodMa}), but $R_{2/3}$ only benefits from the parametric enhancement ${\tilde m}_b/{\tilde m}_t$. Hence, there are two conditions for which the two-body decay will dominate
over the three-body one:
\begin{itemize}
 \item If the $t \bar t$ threshold is closed one needs a somewhat hierarchy between the 
sbottoms
and the stops,
 \item Decouple all the squarks at very high masses, in this case $R_{2/3}
\propto m_t^2/\mgino^2 [1+ \ln x_t]^2$.
\end{itemize}
However, as stressed in \cite{Gambino:2005eh}, for such very heavy squarks,
the logarithmic corrections have to be resummed and in the case of all squarks
masses being equal and large, the resummation tends to decrease $R_{2/3}$ compared to Eq.~(\ref{Rtwothree}) when the common scalar mass scale $\tilde m \gtrsim 10^6$ GeV. On the other hand, in the bino-like case, $R_{2/3}$ does not benefit from such parametric enhancements. The lack of logarithmic enhancement for the two-body
decay (\textit{i.e} the higgsino-like neutralino is not allowed in gluino
decays) implies that the three-body will generally dominate instead, if available. 
\par\noi 
It is worth adding that in the pure higgsino case, it is rather another three-body decay
which rival with the gluino loop decay, namely $\gino \ra t \bar{b}/\bar{t} b \charg{1}$. 
This is due to the fact that the neutral and charged higgsinos are almost degenerated in 
mass, and this decay will dominate only once the $t \bar{b}/\bar{t} b$ threshold is 
crossed, \textit{i.e} $\mgino - \mneut{1} \simeq \mgino - \mcharg{1} > m_t + m_b$. These 
three-body 
decays do not suffer from a hierarchical third generation squark spectrum since they are 
mediated by stops. In the pure bino case this decay will be effective at larger $\Delta M$ 
since the lightest chargino $\charg{1}$ is much heavier than the bino-like LSP.
\par\noindent
In order to illustrate to which extent the radiative gluino decay can dominate over the 
other decay modes, without assuming any specific SUSY-breaking pattern, we performed a numerical scan within the pMSSM parameter space using \texttt{SUSY-HIT} \cite{Djouadi:2006bz}. In this numerical investigation we set all soft masses 
(sleptons and squarks) to 5 TeV and trilinear couplings $A_i$ to zero except for the 
third generation squarks for which we chose the following range:
\begin{itemize}
 \item $M_{\tilde t_R} = 1$ TeV, $M_{{\tilde Q}_3} = 2$ TeV, $A_b=0$
 \item 1 TeV $<$ $M_{\tilde b_R}$ $<$ 2 TeV
 \item $-2$ TeV $<$ $A_t$ $<$ 2 TeV
\end{itemize}\noi 
The parameters related to the gaugino and Higgs sector were set to
\begin{itemize}
 \item 400 GeV $<$ $M_1,\mu$ $<$ 800 GeV
 \item $M_2 = 2$ TeV, $M_3 = 600$ GeV, $\tan\beta = 10$
\end{itemize}\noindent
The scalar mass scale chosen is sufficiently low such that we ensure that the gluino would decay in the detector. We picked up only points giving the SM-like Higgs mass in the range $122\,\mbox{GeV} < m_h < 128\,\mbox{GeV}$ and the resulting gluino mass is driven around $\mgino \simeq 810$ GeV due to radiative corrections. We ensured that $\mgino > \mneut{1}$ such that the lightest neutralino is always the LSP. We divided 
the surviving points of the scan into three different regions, depending on the nature of 
the lightest neutralino, whether it is mostly higgsino-like, bino-like or a mixed 
bino-higgsino. The 
higgsino-like (bino-like) neutralinos correspond to a higgsino (bino) fraction greater than 90\%. 
The remaining points are an admixture of bino-higgsino like neutralinos which we 
dubbed as ``mixed''.
\par
\begin{figure}[h]
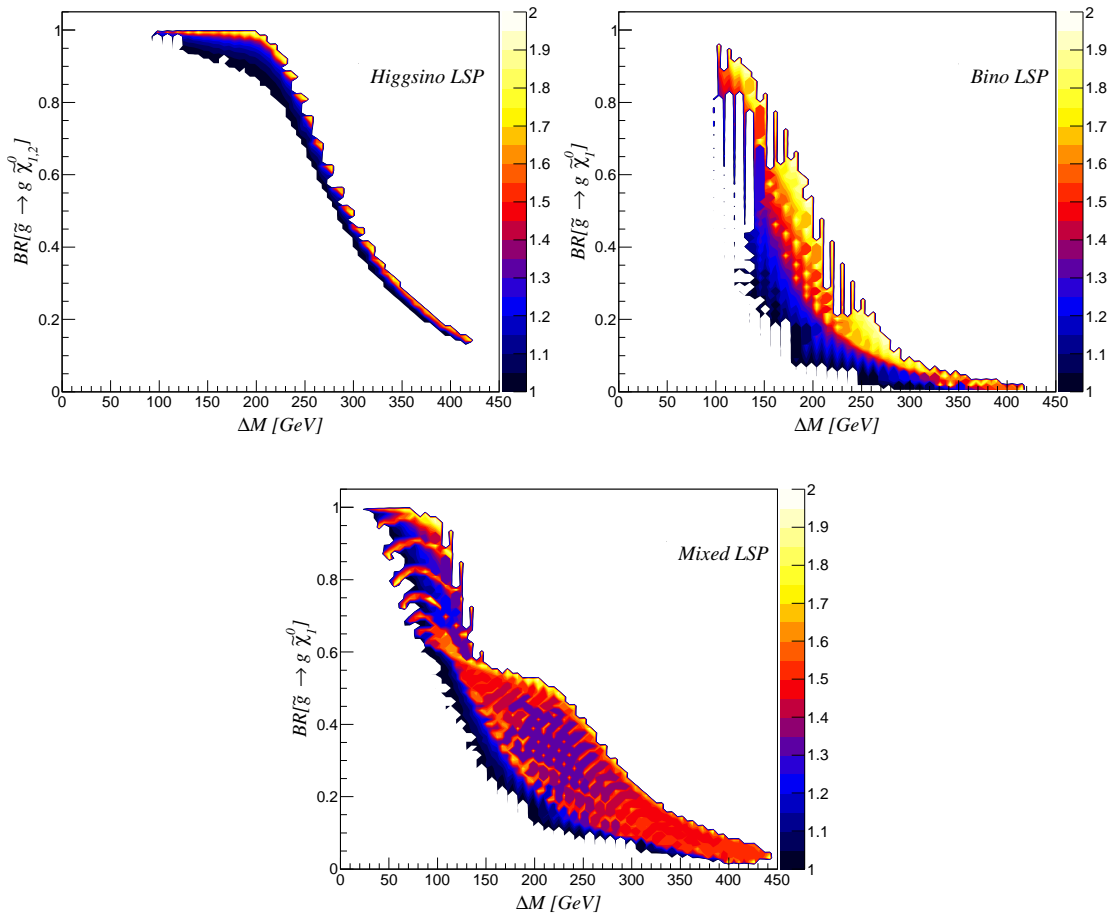

\begin{center}
 \epsfig{file=DeltaMvsBRvsMsquarks_hino,clip=,width=.44\textwidth}
 \epsfig{file=DeltaMvsBRvsMsquarks_bino,clip=,width=.44\textwidth}
 \epsfig{file=DeltaMvsBRvsMsquarks_mixed,clip=,width=.44\textwidth}
\end{center}\caption{\em \label{scan2body}Branching ratio of $\gino \ra g \neut{1}$ 
in terms of $\Delta M = \mgino-\mneut{1}$ for a higgsino-like (top,left), bino-like 
(top,right) and mixed bino-higgsino (bottom) neutralino. In the higgsino-like case the 
branching ratio of the radiative decay $\gino \ra g \neut{2}$ is also included. The colour shading indicates the ratio $m_{{\tilde b}_1}/m_{{\tilde t}_1}$.}
\end{figure}

\begin{figure}[h]
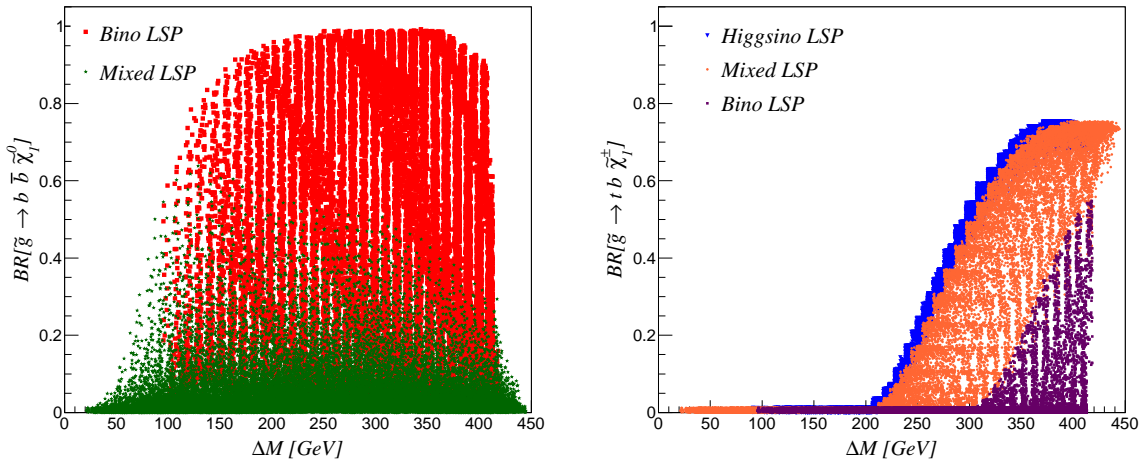

\begin{center}
 \epsfig{file=DeltaMvsBR_3bodybb,clip=,width=.47\textwidth}
 \epsfig{file=DeltaMvsBR_3bodytb,clip=,width=.47\textwidth}
\end{center}\caption{\em \label{BR3body} Branching ratios for the most important 
three-body decay modes depending on the composition of the LSP. The left panel 
corresponds to the channel $\gino \ra b\bar{b}\neut{1}$, where red squares correspond to 
the case where the LSP is bino-like and green stars when it is a mixed bino-higgsino. The pure higgsino case is not shown as it is negligible. The right panel displays the $\gino \ra t \bar{b}/\bar{t} b  \charg{1}$ for each $\neut{1}$ composition, blue diamonds points corresponds to situations where the 
LSP is higgsino-like, orange points and when it is mixed magenta squares.}
\end{figure}

\noi 
We display in Fig.~\ref{scan2body} the branching ratio of the two 
body decay of the gluino into a gluon plus the lightest supersymmetric particle (LSP) in 
terms of the mass difference $\Delta M = \mgino-\mneut{1}$. Given the nature of the 
spectrum, only the radiative decays $\gino \ra g  \neut{i}$ and the three-body modes into 
heavy quarks ($\gino \ra b \bar{b}/t \bar{t}\neut{i}$ and $\gino \ra t \bar{b}/\bar{t} b 
\charg{1}$) are allowed. 
\noi
The colour shading correspond to the level of hierarchy between the lightest stop 
$\stopo$ and lightest sbottom $\sboto$ given by the ratio $m_{\sboto}/m_{\stopo}$. It is 
worth noting that in the higgsino-like case we also included the two-body decay of 
the gluino to the next-to-lightest neutralino $\gino \ra g  \neut{2}$ since the mass 
difference between the $\neut{1}$ and $\neut{2}$ is at most $\simeq 15$ GeV. For such a small mass difference the next-to-lightest neutralino $\neut{2}$ would decay into the LSP and soft leptons or quarks 
originating from the off-shell Z boson decay in $\neut{2}\ra Z^*\neut{1}$. Hence the channel $\gino \ra g \neut{2}$ would leave the same experimental signature of jets + MET in the detector as 
$\gino \ra g  \neut{1}$. Glancing at Fig.~\ref{scan2body} it 
is clear that largest 
branching fractions for the radiative decay are obtained when the mass difference $\Delta 
M$ is small. As one could expect from the approximate formulas in 
Eqs.~(\ref{twobodybino}),(\ref{twobodyhiggino}), the branching ratio of the two body 
decays 
$\gino \ra g \higgino_{1,2}$ (top, left panel of Fig.~\ref{scan2body}) gives the largest 
branching fractions. This is due to the fact that it depends on the fourth power of the 
top mass, unlike the bino-like and mixed cases, see 
Eqs.~(\ref{twobodybino}),(\ref{twobodyhiggino}). In addition, the dependence on ratio 
$m_{\sboto}/m_{\stopo}$ of the loop-induced decay is much more pronounced in the 
bino-like 
and mixed case than in the higgsino case. It is particularly true for the bino case (top, 
right panel) where largest branching fraction are only obtained when the hierarchy between 
the sbottoms and stops is significant. Moreover, the two-body decay into higgsino-like 
neutralinos dominate even for quite large $\Delta M$: the radiative decay starts to 
fall off at $\Delta M \simeq 200$ GeV, while for the bino-like and mixed cases they 
already decline at $\Delta M \simeq 100$ GeV. The drops of the loop-induced decays can be 
understood from looking at Fig.~\ref{BR3body}. In this figure are displayed the branching 
fractions of the most important three-body decays for each nature of the LSP. The left 
panel features the branching ratio $\mbox{BR}(\gino \ra b \bar b \neut{1})$ for 
bino-like or mixed neutralinos. The higgsino-like LSP case is not shown since it is 
negligible and buried in the two others. The right panel presents the branching fraction 
of the gluino decay into $t \bar{b}/\bar{t} b$ final states for all three LSP natures 
considered. In the higgsino case (blue diamonds in the right panel of 
Fig.~\ref{BR3body}), a comparison with the top, left panel of Fig.~\ref{scan2body} makes 
it 
clear that as soon as the decay into $t \bar b/\bar{t} b$ is kinematically allowed,
the branching fraction $\mbox{BR}(\gino \ra g \neut{1,2})$ decreases fast (for 
$\Delta M \gtrsim 200$ GeV) and the three-body mode $\gino \ra t \bar{b}/\bar{t} b  
\charg{1}$ starts to take over. However, thanks to the logarithmic enhancement, the 
loop-decay of the gluino to neutral higgsinos still persists at large $\Delta M$ and is 
of order $10-20\%$.
\noi
The rationale for this decay mode to take over in the higgsino case is that, as soon as 
the threshold is open, it does not suffer from large suppression factors. Indeed, this 
decay is also mediated by stop exchange, as the loop decay is, and not suppressed by 
small quark masses. In the pure bino case, $\mbox{BR}(\gino \ra t \bar{b}/\bar{t} b 
\charg{1})$ is suppressed until larger $\Delta M$. This is due to 
the parameter range we chose for the scan. Indeed, for this decay mode to be open one 
needs a significant mass splitting between the gluino and the lightest chargino, 
typically $\mgino - \mcharg{1} \gtrsim m_t + m_b$. In our scan $ \mcharg{1}$ is 
driven by the value of the parameter $\mu$, which must still be quite large to 
keep the LSP bino-like. Therefore one needs to reach a significantly larger mass 
splitting $\Delta M$ to fulfil these requirements, compared to the pure higgsino 
case where $\mneut{1} \simeq \mcharg{1} \simeq \mu$. Instead, the three-body decay mode 
$\gino \ra b\bar{b}\neut{1}$ quickly dominates over $ \gino \ra g \neut{1}$, already from 
$\Delta M \simeq 100$ GeV. This occurs even when the ratio $m_{{\tilde b}_1}/m_{{\tilde 
t}_1}$ is large. Again, the underlying argument is that the gluino loop decay to a 
bino-like neutralino does not profit from any enhancement, apart from phase space 
considerations. For a mixed neutralino, when $\Delta M$ increases, 
the suppression of the radiative decay is counter-balanced by both $\gino \ra 
b\bar{b}\neut{1}$ and  $\gino\ra t \bar{b}/\bar{t} b \charg{1}$.  
\par\noindent
Having identified in which region of parameter space of the pMSSM the radiative decay can 
dominate, we now turn on to which extent it can be constrained by existing ATLAS and CMS 
8 TeV analyses.

\section{Constraining the gluino loop decay from existing 8 TeV LHC analyses} 
\label{run1}
It is expected that the typical signature of the radiative gluino decay will lead to
a preponderance of jets. If favoured, when the mass difference $\Delta M = \mgino
- \mneut{}$ is small, the final state jets are typically as soft as those
originating from the parton shower and in turn the gluino decay gets buried
into the huge QCD background. To increase the sensitivity one can resort to,
similarly to what is done concerning Dark Matter or compressed spectra
searches, trigger on a hard initial-state-radiation (ISR) jet, the so-called monojet searches. 
\par\noi 
Since the radiative gluino decay has not been fully investigated at the LHC, we wish to fill 
this gap, using in particular monojet analyses to be sensitive to it. It is worth to note 
that the multijet analysis performed in \cite{Aad:2015iea} has indeed probed a small mass 
$\Delta M$. Nevertheless, the interpretation has been done in a different context 
from ours, where the first and second generation squarks are not too heavy and therefore 
the three-body decay into light-flavour quarks $\gino \ra q \bar{q} \neut{1}$ is not 
suppressed. Thus, one can probe small mass splitting between the gluino and the LSP in 
this case. In \cite{Aad:2015iea}, the direct decay of the gluino to a gluon 
and a neutralino was also used to constrain $\mgino$ and $\mneut{1}$, but based only on a 
fully hadronic search. Only numbers for particular values of $\mgino$ and $\mneut{1}$ 
were presented. For a massless neutralino, gluino masses below 1250 GeV can be 
excluded, and a lower mass limit of $\mneut{1} = 550$ GeV corresponding to $\mgino = 850$ 
GeV is quoted, but no full coverage of the $\mgino-\mneut{1}$ mass plane has been 
presented. In \cite{Arbey:2015hca}, an exclusion contour in the $\mgino-\mneut{1}$ mass 
plane has been derived using monojet searches, using the process $\gino \ra q 
\bar{q} \neut{1}$ assuming 100\% branching fraction, among other scenarios investigated. 
However, as we saw in the previous section, the dominance of this channel strongly depends 
on the squark spectrum, which is not detailed in \cite{Arbey:2015hca}, except that the 
stops and sbottoms are assumed to be light. We saw in the previous section that the 
radiative decay is favoured when light stops are present and therefore it is not clear if 
the assumption that $\gino \ra q \bar{q} \neut{1}$ dominates holds.
\par\noi
In the context where the first and second generation of squarks are degenerate, bounds 
on their masses are pushed beyond the TeV scale.  In the scenario we are focussing on, namely with
decoupled first and second generations with somewhat lighter third generation, this effort is 
complementary to the official experimental analyses and worth to investigate. Achieving such a task then requires to recast the already existing ATLAS and CMS analyses since we do not have access to the raw data of 
both collaborations. The experimental signature for the gluino radiative decay is rather 
simple: 
it only consists of jets (possibly very soft) and missing transverse energy (MET). A 
characteristic feature of scenarios where the two-body radiative gluino decay dominate is 
typically the ones which are difficult to probe in a collider environment, namely ones where the mass difference 
between the mother and daughter particles is small (\textit{i.e} compressed spectra). 
These scenarios are difficult to explore since they lead to final states with less 
energetic states or leptons, thus reducing the detection efficiency and signal acceptance. 
This may be part of the reasons why this decay pattern has not been fully considered as 
worth to investigate first in the past, as a priori its prospects are not very 
enthusiasming. In addition, in 
typical mSUGRA/CMSSM scenarios, this branching ratio is quite small since the mass 
splitting between the gluino of the LSP is generically quite large. However, the advent 
of monojet searches to precisely increase the sensitivity to compressed spectra has been 
a game changer in this respect. As noted in \cite{Sato:2012xf}, when the ratio of the 
Higgs doublet vev's $\tan \beta \simeq {\cal O}(1)$, a 125 GeV Higgs implies a 
non-negligible branching fraction of the radiative modes for the gluino decay. Moreover it 
is well-motivated in SUSY scenarios where the gluino is the 
next-to-lightest supersymmetric particle (NLSP), the so-called GNLSP models. Examining this process at the LHC
can also put constraints on the viability of these models in solving the Dark Matter 
problem. Indeed, in such models, the right amount of relic density is obtained through 
important coannihilation processes between the LSP and the gluino and their efficiency is 
mainly driven by $\Delta M$ see for example, \cite{Ellis:2015vaa} and references therein. 
Hence, a full quantitative investigation of the radiative gluino at the LHC seems timely. 
\par\noi 
The experimental signature of the radiative decay can be categorised in 1, 2 and 3 jets 
bins plus MET. Unfortunately there exists no published official analysis covering all such 
categories. In this purpose, and to cover all jet bins, we reimplemented within the 
\texttt{MadAnalysis5}\cite{Conte:2012fm,Conte:2014zja} framework the following three analyses:
\begin{itemize}
 \item ATLAS-SUSY-2013-21 \cite{Aad:2014nra}: this analysis targets the stop decay $\stopo \ra c 
+ \neut{1}$ using a monojet analysis. It has also been used to probe the small 
squark-neutralino mass difference using $\tilde{q} \ra q + \neut{1}$, see 
\cite{Aad:2015iea},
 \item ATLAS-SUSY-2013-02 \cite{Aad:2014wea}: search for squarks and gluinos in final states 
using high-$p_T$ jets with no leptons, 2--6 jets and large MET,
 \item CMS-SUS-13-012 \cite{Chatrchyan:2014lfa}: search for new physics in multijet events with large 
MET divided into three jet multiplicity categories (3--5, 6--7, and 8 jets),
\end{itemize}\noi 
in order to work out the 8 TeV LHC constraints using 20.3 ${\rm fb}^{-1}$ of collected 
data. The recasted CMS-SUS-13-012 SUSY search was already available in the \texttt{MadAnalysis5} Public 
Analysis Database \cite{Dumont:2014tja} and can be found at \cite{cms:13012}. 
The two remaining searches were implemented and validated for the present work and are now 
available at \cite{atlas:13002,atlas:13021}. For the sake of clarity, we present in the 
Appendix \ref{app_monojet} a comparison 
between the official cut flows and our reimplementations of the ATLAS monojet 
\cite{Aad:2014nra} and multijet \cite{Aad:2014wea} analyses within \texttt{MadAnalysis5} 
(\texttt{MA5}). From Tables~\ref{mono_cutflow},\ref{tab:cutflow1} and \ref{tab:cutflow2} 
the selection cuts of each respective analysis can be read off.
\par\noindent
 For the monojet search, see Table~\ref{mono_cutflow}, one can see that the agreement is 
not very good at the level of the preselection cuts. The discrepancy originates already 
from the $E_{T}^{\rm miss} > 100$ GeV cut applied at the reconstructed level. The MET 
trigger efficiency varies with $E_{T}^{\rm miss}$ and in particular is not 100\% effective 
at $E_{T}^{\rm miss} = 100$ GeV. To reproduce the MET trigger efficiency we parametrised 
the efficiency turn-on curve presented in \cite{mettrigger} coming from the ATLAS 
simulation of the process $pp \rightarrow Z H \rightarrow \nu \bar{\nu} b \bar{b}$, as 
advised after communication with the ATLAS SUSY conveners. We already observe at 
this level a discrepancy which ranges from 16\% to 32\% for the three benchmark points 
between our implementation and the ATLAS result. Moreover we cannot reproduce the 
Trigger, Event Cleaning, and Bad Jet veto efficiencies, as we do not have access to this 
information in official documentation. However, once we enter the signal regions, the 
relative efficiencies of our cuts are comparable with the official ones and at the level 
of final number of events we observe an agreement between 15-25\%, which is in the 
ballpark of expected accuracies from fast-simulation. More details about the validation 
of the recasting procedure can be found at \cite{MADPAD}. 
\par\noi 
For the ATLAS multijet analysis \cite{Aad:2014wea} a cut flow comparison is also 
presented 
in Tables~\ref{tab:cutflow1} and \ref{tab:cutflow2}. One can see that the agreement 
between the recasted analysis and the official is fairly good. The CMS multijet 
implementation and validation was already available on the PAD \cite{Dumont:2014tja} 
before the initiation of this work and all details were presented there.
\par\noi
As official results are not provided in terms of the gluino branching ratios and the 
radiative mode has not been fully investigated so far, we derived limits on this decay 
pattern within a Simplified Model approach. In the case we focussed only the gluino and 
LSP are relevant in the mass spectrum and we assume $\mbox{BR}(\gino \ra g \neut{1}) = 
100 \%$. The signal generation was done using 
\texttt{MADGRAPH5}\cite{Alwall:2011uj,Alwall:2014hca} 
using the CTEQQ6L1 parton distribution functions \cite{Pumplin:2002vw}, as
\begin{equation}
 p p \ra \gino \gino + X \ra j j + E_{T}^{\rm miss}
\end{equation}\noi 
where $X = 0,1$ is the number of generated jets in the signal sample. The 
\texttt{MADGRAPH} samples are produced using the AUET2B 
tune\cite{ATL-PHYS-PUB-2011-009,ATL-PHYS-PUB-2011-014}\footnote{It is worth to note that 
care should be taken when using recent versions of \texttt{MadGraph5} together with the 
AUET2B tune. Indeed this tune was designed from a Leading-Order (LO) Matrix Element + 
Parton Shower (MEPS) merging. Since \texttt{MadGraph5.2}, the matching MEPS is done at 
next-to-leading order and in the same fashion as in MC\@NLO \cite{Frixione:2002ik}. 
Therefore if one uses this tune together with a version 5.2 or higher of 
\texttt{MadGraph}, the interpretation of the results should be taken with great care. 
Indeed, we observed significant discrepancies in reproducing the official exclusion 
curves between using a recent version of \texttt{MadGraph} or the one actually adopted in 
the official analyses in \cite{Aad:2014nra,Aad:2014wea}. That is why, for each analysis, 
we stick as much as possible to the Monte-Carlo configurations described in the 
experimental analyses. Thus, each signal sample has been generated with a different 
\texttt{MadGraph} version for each analysis we recasted.}. The MLM \cite{Mangano:2006rw} 
matching scheme is used,  and a \texttt{MADGRAPH} $k_T$ measure cut-off and a
\texttt{PYTHIA}\cite{Sjostrand:2006za} jet measure cut-off both set to 0.25 times the mass scale of the SUSY 
particles produced in the hard process. Hadronisation is performed using 
\texttt{PYTHIA-6.426}\cite{Sjostrand:2006za} as incorporated in the \texttt{MADGRAPH-5} package and detector 
effects are simulated with the 'MA5tune' of \texttt{DELPHES-3}\cite{deFavereau:2013fsa} as 
described in \cite{Conte:2014zja}. The number of events are then normalised to the  
correct luminosity after including cross sections at the next-to-leading order and 
next-to-leading logarithm accuracy \cite{Kramer:2012bx}, as tabulated by the LHC SUSY 
Cross Section 
Working Group \cite{8tevxs_susy}
\par
\begin{figure}[htbp]
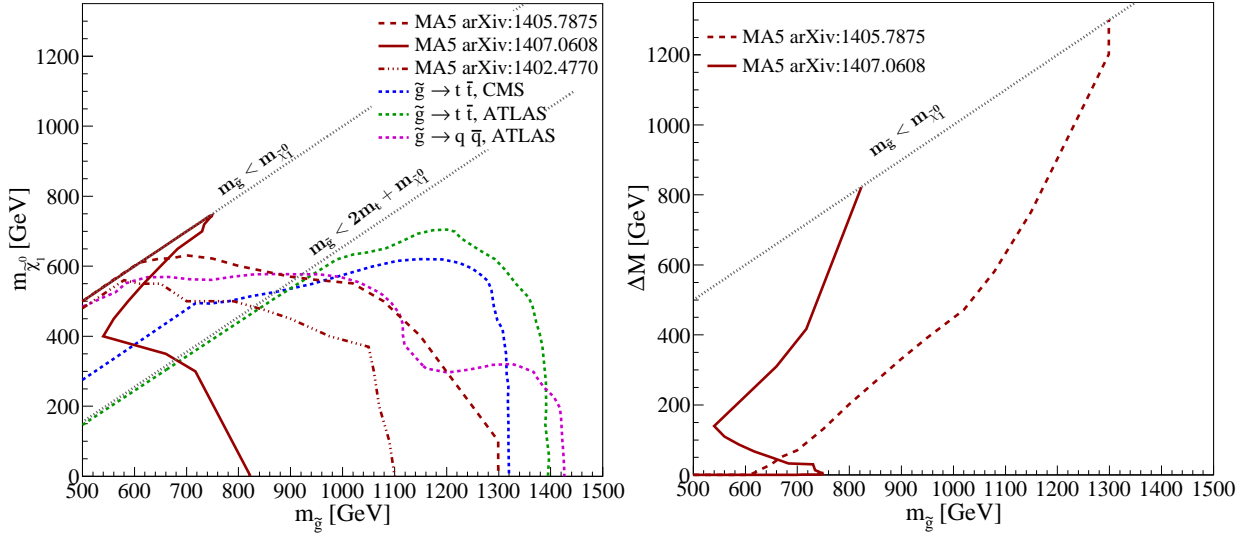

 \begin{center}
  \epsfig{file=MA5_exclusion_fin.eps,clip=,width=.49\textwidth}
  \epsfig{file=exclusion_DeltaM,clip=,width=.49\textwidth}
 \end{center}\caption{\label{8TeVlimits} \em Left panel: 95\% CL exclusion 
contours for the radiative gluino decay simplified topology. The solid red line 
corresponds to the mass limits obtained from the \texttt{MA5} recasted
ATLAS monojet analysis \cite{atlas:13021}, the red broken line from the \texttt{MA5} 
recasted ATLAS multijet search \cite{atlas:13002} and the dashed-dotted line from the 
\texttt{MA5} recasted CMS multijet analysis \cite{cms:13012}.
For comparison, the official 95\% CL exclusion lines for the $\gino \ra t \bar{t} \neut{1}$ SMS from ATLAS (green dashed line) \cite{Aad:2015iea} and CMS \cite{Chatrchyan:2013iqa}(blue dashed line) 
as well as for the $\gino \ra q \bar{q} \neut{1}$ SMS from ATLAS \cite{Aad:2015iea}(purple dashed line) are also shown. Right panel: 95\% CL exclusion contours in the plane $\mgino$ versus $\Delta M$ where $\Delta M = \mgino-\mneut{1}$. Only the exclusion contours from the \texttt{MA5} recasted monojet and multijet analyses are given since they provide the best sensitivity in the low $\Delta M$ region.
}
\end{figure}

\noi 
For the statistical interpretation, we make use of the module \texttt{exclusion\_CLs.py} 
provided in the \texttt{MadAnalysis5} recasting tools. Given the number of 
signal, expected and observed background events, together with the background 
uncertainty (both directly taken from the experimental publications), 
\texttt{exclusion\_CLs.py} determines the most sensitive signal region (SR), the 
exclusion 
confidence level using the $\mbox{CL}_s$ prescription from the most sensitive SR, and the 
nominal cross section 
$\sigma_{95}$ that is excluded at 95\% CL. The simplified model parameter space is simply 
scanned by varying $\mgino$ and $\mneut{1}$ until a 95\% CL excluded number of BSM 
signal events is reached. The derived limits in the $(\mgino,\mneut{1}$) plane are 
presented in Fig.~\ref{8TeVlimits}.
\noi 
In the left panel of Fig.~\ref{8TeVlimits} the red lines are the 95\% CL exclusion bounds using the recasted analysis\footnote{Note that one could have used, for the monojet analysis, the acceptance times efficiency 
tables as provided on the \texttt{Twiki} page of \cite{Aad:2014nra}. We chose to 
recast anyway this analysis to enrich the PAD \cite{Dumont:2014tja}.}. The results of the 
three recasted analyses (solid line : ATLAS 
monojet \cite{atlas:13021}, broken line: ATLAS multijet \cite{atlas:13002} and 
dashed-dotted: CMS 
multijet \cite{cms:13012}) are compared against three official published results: in 
green and blue are displayed the 95\% CL limits obtained from the simplified topology 
$\gino \ra t \bar{t} \neut{1}$ from ATLAS \cite{Aad:2015iea} for the former and from CMS 
for the latter \cite{Chatrchyan:2013iqa}. One immediately observes that the official SMS 
interpretations are limited when the $t \bar{t}$ threshold of the gluino three-body decay 
is closed. The CMS results pushes the limit a little bit 
above the $t \bar{t}$ threshold limit using off-shell tops, as compared to ATLAS. The 
purple broken line corresponds to the ATLAS limits in the $(\mgino,\mneut{1}$) mass plane 
for the $\gino \ra q \bar{q} 
\neut{1}$ simplified topology using the multijet search \cite{Aad:2015iea}. In the corresponding model 
investigated all squarks are decoupled and since the final states quarks are massless in 
this simplified approach, a small $\Delta M$ can be probed. In our scenario only the 
light-flavour squarks are completely decoupled from the low-energy spectrum and therefore 
our results are complementary to these. 
\par\noi 
A word of caution is in order before discussing and interpreting our results. Indeed, we 
do not provide an estimation of the theoretical uncertainties on both the signal cross 
section and acceptance. This certainly limits our ability to provide an unambiguous 
statement on which portion of the parameter space in the $(\mgino,\mneut{1}$) mass plane 
is excluded or not, but nevertheless provide a very interesting estimation of the sensitivity 
and mass reach of the 8 TeV analyses we reimplemented. Nevertheless, comparing with the 
existing literature enabled us to estimate the correctness of our results. First of all, 
comparing with Fig.~4 of \cite{Arbey:2015hca}, we see that the shape of the monojet 
exclusion curve is similar and its maximum $(\mgino,\mneut{1})$ excluded values lie at 
$(\mgino,\mneut{1})\simeq (700,700)$ GeV, which is quite similar to our own maximum which 
sits at $(\mgino,\mneut{1})\simeq (750,750)$. In addition, the kink appearing for 
$(\mgino,\mneut{1}) \simeq (540,400)$ GeV is also observed in other interpretations of 
monojet analyses (see \cite{Arbey:2015hca,Aad:2015iea}). The underlying explanation of the 
kink comes from the fact that the monojet analysis is not, strictly speaking, based on a 
single hard jet. Indeed, the monojet selection cuts in \cite{Aad:2014nra} actually allow 
a number of jets $N_{\rm jets} \leq 3$ with $p_T > 30$ GeV and $|\eta| < 2.8$. Thus the 
appearance of the kink is due to the fact that, starting from this point which possesses 
a 
sufficient $\Delta M$, additional jets fulfilling the above requirement populate the 
SRs, 
thereby enhancing the sensitivity.  Next, comparing with the numbers given in Fig.~21 in 
\cite{Aad:2015iea}, for a massless neutralino, gluino masses up to $\mgino = 1250$ GeV are 
excluded, whereas we exclude up to $\mgino = 1300$ GeV (see the broken red line in the 
left panel of Fig.~\ref{8TeVlimits}). A lower mass limit of $\mneut{1}=550$ GeV for 
$\mgino = 850$ GeV is also quoted to be compared with $\mneut{1}\simeq 600$ GeV, $\mgino 
= 850$ GeV on our side. Therefore our results are in fairly good agreement with the 
existing litterature. Last, there has been numerous studies, mostly in
the context of Simplified Dark Matter models (see {\it e.g} \cite{Fox:2011fx,Fox:2011pm}),
showing that the mono-jet searches can be sensitive to the decay width of the
intermediate particles. However, since in our case the gluino decays through
only one decay mode, its total width $\Gamma$ is suppressed and $\Gamma/\mgino
\ll 1$, such that we can use safely the narrow-width approximation implicitly
assumed in our SMS approach.
\par\noi 
Provided this important clarification, let us now discuss our results. As expected, the 
monojet analysis is the most sensitive search to probe the scenario where the gluino 
and neutralino are almost degenerate. Mass-degenerate gluino and LSP are excluded up to 
750 GeV. As soon as $\Delta M$ increases, the efficiency of the monojet analysis quickly 
drops. This can be seen on the right panel of Fig.~\ref{8TeVlimits}, where we display the 
two most sensitive recasted analyses in the small $\Delta M$ region. The colour coding is 
the same as in the left panel. The monojet analysis is the most effective in constraining 
a compressed gluino-neutralino spectrum up to $\Delta M \simeq 50$ GeV.
\par\noi 
For larger $\Delta M$ the ATLAS multijet analysis becomes the most sensitive one in 
constraining the radiative decay. At low $\mgino$ the signal region dubbed as 2jl (``2 
loose jets '', more details about the SR definition can be found in the official analysis 
\cite{Aad:2014wea}) drives the exclusion curve, at moderate $\mgino$ it is driven by SR 
2jl and 3jm (``3 medium jets'') and at large $\mgino$ the 4jl (``4 loose jets'') SR 
dominates. Within this analysis gluino masses up to $\mgino \simeq 1300$ GeV for a 
massless neutralino are excluded. It appears that the CMS multijet analysis is less 
powerful than the two ATLAS ones. There are two reasons for that, one is that the jet 
binning only starts from 3 jet bins and the second is that the search was designed as a 
discovery one, therefore its constraining power is less effective. Another comment is 
worth to be mentioned here: below the line where the $t \bar{t}$ threshold is open, 
interpreting the 
exclusion contour is highly debatable since we expect the three-body decays to take over. 
Therefore below this line our results should be taken with caution and the official 
results focussing on the three-body decay $\gino \ra t \bar{t}$ are more realistic.
\par\noindent
In the light of these results, it would be interesting to perform a combination of at least the two ATLAS analyses to get a more realistic exclusion curve. However, this goes beyond the scope of this paper since we do not have the necessary information at our disposal to perform the combination.

\section{Prospects at the 13 TeV LHC}

We address the discovery and exclusion reach of the radiative 
gluino decay at 13 TeV in this section.
Since the radiative decay is dominant with a small mass gap 
between $\tilde{g}$ and $\widetilde{\chi}_{1,2}^{0}$, we design a special 
set of cuts optimized to investigate this case.
  To simplify matters we select benchmark points with
 a $\tilde{g}-\widetilde{\chi}_{1,2}^{0}$ SMS in mind, where the decay
 $\tilde{g}\to g\widetilde{\chi}_{1,2}^{0}$  is 100 $\%$. As we  already observed
 in section \ref{pheno}, such 
an assumption is well motivated from theoretical considerations.  We choose three representative benchmark points(BP)  with nearly degenerate masses for $\rm \tilde{g}$ and $\rm \widetilde{\chi}_{1}^0$ . 
\par\noi   
To achieve this we decouple the rest of the spectrum from this set, i.e we set the  three generation of squarks and the sleptons to 3 TeV.
   Since a higgsino like scenario is favourable for the 
radiative decay,
  we set $M_1$, $M_2$ to 2 TeV, while $\mu$ is kept relatively 
light so that $\widetilde{\chi}_{1}^{0}, \widetilde{\chi}_{2}^{0}$ are Higgsino like.
 For all  benchmark 
points, the mass gap $\Delta m = m_{\tilde{g}} - m_{\widetilde{\chi}_1^{0}}$ is fixed to around 10 GeV.
 The availability of phase space ensures that both  
$\tilde{g}\to g \widetilde{\chi}_{1}^{0}$ and $\tilde{g}\to g \widetilde{\chi}_{2}^{0}$ are equally 
important. However, since $\widetilde{\chi}_{1}^{0}$ and $\widetilde{\chi}_{2}^{0}$ are 
almost degenerate,
 the decay products of $\widetilde{\chi}_{2}^{0}$ are soft, and are most likely to go undetected.
\par\noi
   The relevant masses and the branching ratios (BR) are tabulated in Table \ref{tab-spec}.
The benchmark points were generated using \texttt{SUSYHIT} \cite{Djouadi:2006bz}. 
 All the benchmark points satisfy the constraints derived in the previous section.
\begin{table}[ht!]
\small
\begin{center}
\tabulinesep=1.2mm
\begin{tabu}{|c|c|c|c|}
\hline
  &  BP1 & BP2 & BP3   \\
\hline
$M_3$ & 730 & 820 & 1100  \\
$\mu$ &985  & 1180  & 1370   \\
\hline
 $\mgino$  &1005  & 1205 & 1405     \\
\hline
$\mneut{1}$  & 995 & 1195 & 1395    \\
\hline
$\mneut{2}$  & 1004 & 1201 & 1398     \\
\hline
$\mbox{BR}(\gino \to g \, \neut{1}) (\%)$ 
& 99 & 53 & 47    \\
\hline
$\mbox{BR}(\gino \to g \, \neut{2}) (\%)$ 
& 1 & 45 & 47
\\
\hline
\end{tabu} 
\caption{ \it{ The masses and branching ratios for the three benchmark points 
used for the collider analysis.}
\label{tab-spec}}
\end{center}
\end{table}   

\par\noi
 Since we focus on the nearly mass degenerate $\tilde{g}-\widetilde{\chi}_{1}^{0}$ scenario, the conventional
 search strategy is to look for a monojet + missing energy signature, where the hard
 ISR monojet recoils against the missing energy. However, at 13 TeV 
the number of QCD radiation jets from the gluino can be fairly large and carry large 
transverse
momenta. We argue in this section, that a dijet + missing energy scenario (from ISR/FSR) is better suited 
compared to the traditional monojet + missing energy search strategy. 
We argue that the number of hard ISR jets can be significantly large at
13 TeV LHC, and 
including a second ISR jet helps considerably in improving the signal to background ratio.

\begin{figure}[h]
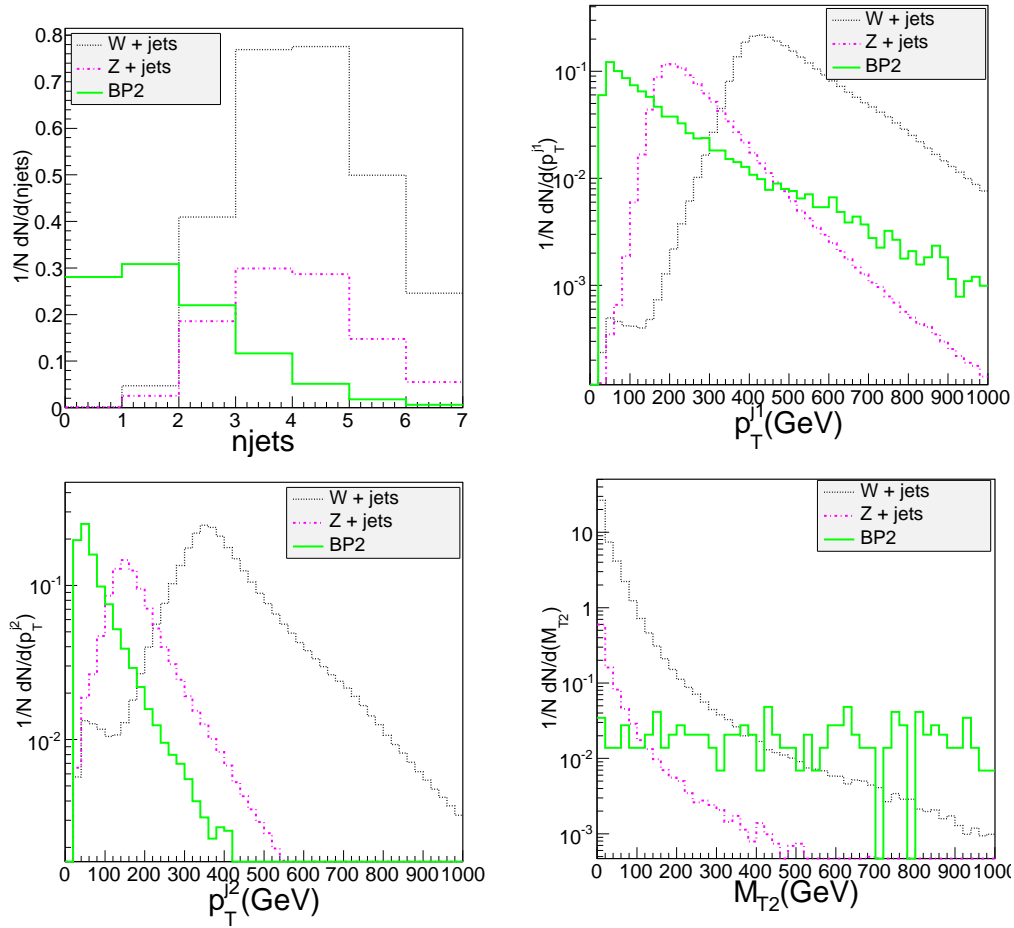

 \begin{center}
  \epsfig{file=njets,clip=,width=.4\textwidth}
  \epsfig{file=ptjet1,clip=,width=.4\textwidth}
  \epsfig{file=ptjet2,clip=,width=.4\textwidth}
  \epsfig{file=mt2,clip=,width=.4\textwidth}
        \end{center}\caption{\em \label{fig:ptjet}\it{ The distribution of the number of njets(top left),
 and the transverse momentum of the leading jet($p_{T}^{j_1}$,top right) are presented in the top panel,
 while the transverse momentum of the second leading jet($p_{T}^{j_2}$, bottom left) and the $\rm M_{T2}$(bottom right)
 distribution is plotted in the lower panel for the signal benchmark point BP2 and the W/Z + jets backgrounds. The $M_{T2}$ distribution is subject to all other cuts, as described in the text. The figures are normalized to unity.}}
\end{figure}\noi

The principal backgrounds to this process are listed below, 

\begin{itemize}
\item {\bf{QCD}}: The QCD dijet events are the largest
source of backgrounds in terms of cross section ($\rm \sim~10 ^{8} pb$).
A significant amount of missing energy is also 
likely to be present from detector mis-measurements. 

\item {\bf{{$\rm t\bar{t}$} + jets}}: The multijet + missing energy topology 
is fairly common for all hadronic top decays. It is possible that the leptons 
and jets are missed which can lead to a dijet + missing energy topology. 
 Since the $t\bar{t}$ + jets
 cross section is significantly large at 13 TeV ($\sim \rm 900$ pb), even a small 
efficiency after cuts can result in a significantly large background cross section. 

\item{\bf{Z + jets}}: This constitutes the irreducible part of the background 
when the Z boson decays invisibly to a pair of neutrinos. Since the cross section
 is quite large ($\rm \sim 10^{5} pb $),
 suppressing this background to a reasonable level is of primary requirement. 

\item {\bf{ZZ}}: This background is similar to the Z + jets background, with one of the Z 
boson
 decaying hadronically while the other invisibly. This background is thus also one of the
 irreducible backgrounds for our case.

\item {\bf{W} + jets}: For a leptonic decay of W, a missed lepton would mean a
 topology similar to the signal. This has to be therefore considered
 as a serious background for this topology. 

\item {\bf{WZ}  }: The WZ background can be significantly large
 for a hadronically decaying W and an invisibly decaying Z.    
\end{itemize}
\par\noi
Parton level events for the signal and background are generated using {\texttt{MADGRAPH5.0.7}}\cite{Alwall:2011uj},
 and passed on to {\texttt{PYTHIA6}} \cite{Sjostrand:2006za} for showering and 
hadronisation.  
For the signal additional two partons are generated at the matrix 
element level. For backgrounds like $t\bar{t}$ + jets, W/Z + jets, 
 additional 3 partons are generated in {\texttt{MADGRAPH5}}.
The merging parameter for the matrix element-parton showering matching
 is set to 50 GeV.  Since we anticipate that 
the bulk of the backgrounds are likely to originate from the tail 
of the $p_{T}$ spectrum we simulate W/Z + jets with a hadronic energy $H_{T}$(defined as $\Sigma p_{T}^{j}$) cut 
of 400 GeV at the parton level in {\texttt{MADGRAPH5}}.
 The sample is then passed on to {\texttt{Delphes3}} \cite{deFavereau:2013fsa}
 (specifically the \texttt{MadAnalysis5} tune \cite{Conte:2014zja} ) for detector simulation.
 Jets are reconstructed using {\texttt{Fastjet}} \cite{Cacciari:2011ma} with an
 anti-$k_{T}$\cite{Cacciari:2008gp} algorithm
 and a $p_{T}$ threshold of 30 GeV and $|\eta_{j}|\le  3$.
Background cross sections are calculated using {\texttt{MADGRAPH5}}, while  
the gluino pair production cross section with decoupled sfermions are 
are quoted from the official SUSY cross section working group page \cite{Borschensky:2014cia}. 
 We use the {\texttt{MadAnalysis5}}\cite{Conte:2014zja} framework for the 
  analysis of signal and background.

\par\noi
The following cuts are  applied to maximize the signal to background ratio: 

\begin{itemize}

\item {\bf{ Lepton Veto (C1)}}: Since the signal has no leptons, a lepton veto 
is imposed. Leptons are selected with a with transverse momentum threshold, $p_{T}^{l}\ge 10~ GeV$ and $|\eta|\le 2.5$. Leptons are
isolated by demanding that the scalar sum of the transverse momenta of all visible stable particles within a cone 
$\Delta R=0.2$, does not exceed 10 $\%$ of $p_{T}^{l}$. This ensures that backgrounds
 with leptonic decays of W/Z are suppressed.  

\item {\bf{Jets (C2)}}: We demand exactly two jets in the final state. From the 
plot of number of jets (njets) for the signal and background (top left 
Fig. \ref{fig:ptjet}), we observe that even with a small mass gap for the signal, it is 
possible to obtain up to 3 jets quite easily from QCD radiation. The background, 
especially, for the W+ jets, peaks at 4 jets.  Therefore it is convenient 
to design a search strategy with a dijet signature.        

\item {\bf{ b-jet veto (C3)}}: To suppress the large $t\bar{t}$ background, 
we veto b-jets. This also ensures that backgrounds like $Zb\bar{b}$,     
$ Wb\bar{b}$ are suppressed.  

\item  {\bf {$p_{T}^{j_1,j_2} > 600,200~ GeV$ (C4)}}: From Fig. \ref{fig:ptjet} we note 
that the leading ISR jet can be 
  significantly energetic for the signal (top right, Fig. \ref{fig:ptjet})
 and therefore we impose a hard cut on the transverse momenta of the leading jet. 
Although the second leading jet is not as energetic (bottom left, Fig. \ref {fig:ptjet}), 
the tail of the distribution can go up to about 400 GeV. We therefore 
impose a cut of 200 GeV on the second leading jet. The hard cuts on the leading
jets ensure that we are in a region where the  missing transverse energy that recoil against these hard ISR 
jets is high.

\item {\bf{ $M_{T2}$ (C5)}}: For this study, the variable $M_{T2}$ 
\cite{Barr:2003rg,Barr:2009wu,Lester:1999tx} is defined as, 
\begin{equation*}
M_{T2}(j_1,j_2,\PMET) = min[max{M_{T}(j1,\PMET^{1}),M_{T}(j2,\PMET^{2})}]
\end{equation*} 
where the minimization is performed over all possible partitions of the missing 
transverse momentum,
  $\PMET = \PMET^{1} + \PMET^{2}$,
 and $j_{1}, j_{2}$, 
represent the two leading jets. The transverse mass of the system is defined as
$$
M_{T}^{2}(j,\PMET) = M_{j}^{2} + M_{\chi}^{2} + 2(E_{T}^{j}E_{T}^{\chi}-\vec{p}_{T}^{j}.\vec{p}_{T}^{\chi})
.$$ The value of 
$M_{T}$ is bounded by the parent particle. It can be shown that this holds true 
for $M_{T2}$ as well, \textit{i.e}, after minimization, $M_{T2}$ is bounded by the mass of 
the 
parent particle.
 To suppress SM backgrounds, where the only source of missing energy 
are the neutrinos, the value of the mass of the invisible particle is set to 0.
 SM backgrounds like WW, WZ, ZZ, $t\bar{t}$, are expected to be cut off at the
 parent particle mass, while since the gluino is massive it is expected to go up to
 higher values. Uncorrelated/ISR jets however can significantly
 alter the shape of the $M_{T2}$ distribution \cite{Belanger:2013oka}. For low mass differences,
 the extra ISR jets that dominate the jet configuration
 are not correlated with the $\PMET$, and hence the 
$M_{T2}$ distribution extends well beyond the gluino mass.
In the bottom left panel of  Fig. \ref{fig:ptjet} we present the 
$M_{T2}$ distribution for the signal(BP2) and the W/Z + jets background subject to all the previously imposed cuts. 
We observe that the 
$M_{T2}$ distribution falls sharply for the background, while the signal 
remains fairly flat. We use an optimized cut of 800 GeV to suppress the background. 
\end{itemize}

\par\noi
We summarize our cut flows for signal and background in Table \ref{tab-cutflow}. For 
simplicity, we only document the backgrounds which leave a non negligible 
contribution after all cuts. In Table \ref{tab-cutflow}, the production cross section (C.S)
for signal and background processes are tabulated in the third column in femtobarn. 
The number of simulated events (N) are noted in the 4th column. From the fifth
column onwards the cumulative effect of cuts are presented.     

\begin{table}
\begin{center}
\begin{tabular}{|c|c|c|c|c|c|c|}
\hline
  & $\mgino,\mneut{1}$ & C.S & N  & C1+C2+C3   & C4
& C5      \\
\hline
 & (GeV) & fb   & & & &    \\ 
\hline
BP1 & 1005,999  & 314  &  100K & 134  & 4.2  & 1.6    \\
BP2 & 1205,1195 & 83   &  100K & 46  & 2.3  & 0.6    \\
BP3 & 1405,1395 & 24   &  100K & 12  & 0.7   & 0.21     \\
\hline
Z + jets &  & $1.7\times10^{8}$ & 5M  &783250  & 12395  & 1.7     \\
\hline
W+ jets & & $5.8 \times 10^{8}$ & 5M  &  31921.1  & 7793.1  & 1.0    \\
\hline
\end{tabular}
\caption{\it{The cut flow table for the signal and background 
 for the analysis. For simplicity
we only provide the processes which yielda  non-negligible  background 
contribution. The cross section after each cut is tabulated from column 5.} }
\label{tab-cutflow}
\end{center}
\end{table}

\par\noi
From Table \ref{tab-cutflow}, we observe that the cut on the transverse
momenta of the leading jets (C4), and $M_{T2}$ (C5) are most effective 
in suppressing W/Z+ jets background. The other backgrounds
(not tabulated) like QCD, $t\bar{t}$ + jets, WW,WZ,ZZ
  are reduced to a negligible level after the application of the $M_{T2}$
cut. Although the signal process suffers as well as a result of 
the $M_{T2}$ cut, the signal to background 
ratio is enhanced considerably. After all cuts, the total background cross section
is reduced to 2.7 fb, while the signal cross sections vary from 1.6 fb for benchmark P1 to  0.21 fb in P3. 
In Table \ref{tab-sig}, we estimate the signal 
significance defined as $S/\sqrt{B}$ by keeping the mass gap between the 
gluino and the lightest neutralino fixed at 10 GeV for the benchmark points.
\par\noi
  We estimate that one can observe a signal at 5 $\sigma$ significance at 13 TeV with
30 $\rm fb^{-1}$ luminosity up to a gluino mass of 1 TeV while a gluino mass of 1.2 TeV
 can be excluded at the same luminosity.
  At 100 $\rm fb^{-1}$ luminosity the discovery
 reach goes up to 1.15 TeV, while we can exclude a gluino mass of up to 1.3 TeV.
 For the high luminosity run of LHC (3000 $\rm fb^{-1}$),
 we can discover a gluino mass up to 1.45 TeV,
  excluding about 1.55 TeV.        
\footnote{For the large mass gap scenario, we easily obtain at least 4 jets from the gluino decay and additional QCD radiation jets.
 In this case the improved CMS and ATLAS multijet searches at 13 TeV are expected to be quite sensitive (as we have already observed for the 8 TeV LHC in section \ref{run1}).
 Indeed we find that our search strategy, which targets a dijet + missing energy 
signature, suffers in this case, and we only manage to discover a gluino mass of 1.1 TeV 
for a $\neut{1}$ mass of 1 TeV with 3000 $\rm fb^{-1}$ luminosity. }
\par\noi
We would like to compare our results with the standard monojet strategy at 13 TeV. However, since this is not available yet, a rough estimate could be obtained
by a rescaling of the cross section from the 8 TeV result to the 13 TeV by assuming that 
the acceptance times efficiency remains the same. This has already 
been performed by the authors of \cite{Arbey:2015hca} for the high luminosity
LHC run (3000 $\rm fb^{-1}$). They predict that one can exclude a gluino 
mass of 1.25 TeV with a $\Delta M$ of 10 GeV, although in the context where  $\mbox{BR}(\gino\to q \bar{q}\neut{1})=100\%$.  
 On this evidence the dijet + missing energy strategy devised here 
has a clear advantage, and we advocate the use of this strategy for 13 TeV LHC.    

\begin{table}
\begin{center}
\begin{tabular}{|c|c|c|c||}
\hline
   &  P1  &  P2  &  P3 \\
\hline
$\mgino,\mneut{1}$(GeV)& 1005,999  & 1205,1195  & 1405,1395        \\
\hline
\hline
 $S/\sqrt{B}$(30 $\rm fb^{-1}$)  & 5.3
  & 2.0 & 0.7    \\
\hline
 $S/\sqrt{B}$(100 $\rm fb^{-1}$)  &
 9.7 & 3.7  & 1.27    \\
\hline
$S/\sqrt{B}$(3000 $\rm fb^{-1}$) & 53  & 20  &  7     \\
\hline
\end{tabular}
\caption{\it{ 
 The signal significances($S/\sqrt{B}$) for different
energies and luminosities for the benchmark points.} 
}

\label{tab-sig}
\end{center}
\end{table}

\section{Conclusion}
We have performed an investigation of the MSSM region where the 
gluino-neutralino spectrum is compressed and the loop-induced decay $\gino \ra g 
\neut{1}$ is dominant over three-body decays. For small gluino-neutralino mass 
differences $\Delta M$, this decay mode is a 
sensitive probe of compressed spectra since it does not suffer from kinematical 
suppression factors.  Within the scenario we investigated, namely with decoupled heavy 
scalars but with third generation squarks closer to the electroweak scale, the 
loop-induced decay of the gluino into a pure higgsino-like neutralino dominates over other 
branching fractions until the $tb$ threshold is crossed, when $\Delta M$ increases. This 
occurs whatever the hierarchy between the stops and the sbottoms, albeit larger branching 
fractions are reached when $m_{\sboto} \gg m_{\stopo}$. On the contrary, the dominance of 
the radiative gluino decay into a bino-like neutralino or mixed bino-higgsino one over 
the other decay 
modes strongly depends on this hierarchy and other three-body decay modes start to be 
significant before the opening of the $tb$ kinematical threshold.
\par\noindent
Adopting a Simplified Model approach, where the gluino and the neutralino $\neut{1}$ are 
the only relevant superparticles accessible at the LHC, and $\mbox{BR}(\gino \ra g 
\neut{1})=100\%$, we reinterpreted 
some ATLAS and CMS 8 TeV SUSY searches to derive 95\% CL exclusion contours in the 
$\mgino-\mneut{1}$ plane. This decay process is particularly interesting to probe small 
$\Delta M$. The reinterpretation of the ATLAS and CMS analyses has been performed through 
the \texttt{MadAnalysis5} framework \cite{Conte:2012fm,Conte:2014zja} and publicly 
available on its Public Analysis Database web page \cite{MADPAD}. The recasted monojet 
analysis enables us to illustrate that, had it been used by the experimental 
collaborations to constrain the loop-induced gluino decay, this search is 
sensitive to degenerate gluino-LSP scenarios. In turn, degenerate situations with gluino 
masses up to $\mgino \simeq 750$ GeV could have been excluded. This surpasses the 
official maximal value $\mgino \simeq 600$ GeV reached by the experimental collaborations 
using the process $\gino \ra q \bar{q} \neut{1}$.
\par\noi 
Then, concentrating on a pure di-jet search, we designed a dedicated search strategy to 
discover the gluino through the $\gino \ra g \neut{1}$ channel. For a mass gap fixed at 
$\Delta M = 10$ GeV, a signal at 5 $\sigma$ significance at 13 TeV with 30 
$\mbox{fb}^{-1}$ luminosity could be observed up to a gluino mass $\mgino = 1$ TeV, while 
the discovery reach goes up to 1.2 TeV. At 100 $\mbox{fb}^{-1}$ luminosity the discovery 
reach goes up to 1.15 TeV, while we can exclude up to 1.3 TeV. A high-luminosity LHC 
could discover a gluino mass up to 1.45 TeV while excluding about 1.55 TeV. 
\par\noi 
In addition, our Monte-Carlo investigation indicated an interesting point. Indeed, a 
pure di-jet search could be as sensitive as a monojet search and maybe even better. The 
underlying reason is related to the fact that a heavy gluino radiates a lot of high-$p_T$ 
ISR jets, even for small $\Delta M$. Of course this has to be taken with a grain of salt 
since our investigation was performed only a the Monte-Carlo level and only a full 
detector simulation could give a clear cut answer, but this result is nevertheless worth 
to be pointed out.
\par\noindent
The discovery of a 125 GeV Higgs boson suggests that the scalar superpartners may lie in 
the multi-TeV range. Direct search of such heavy particles is impossible at the 
LHC. If the scalar mass scale $\tilde m$ is much less than $\tilde m \lesssim {\cal 
O}(10^3)$ TeV, the gluino two-body decay is quite important in probing indirectly 
$\tilde m$. In the end, it may well be our only probe at our disposal in the case of 
SUSY scenarios with heavy scalars and compressed SUSY, if the gluino decays promptly in 
the detector. We hope our work will motivate the experimental collaborations to perform a 
dedicated analysis to search for this decay in order to make the most of the LHC 
potential.
\section*{Acknowledgements}
We thank our experimental colleagues Jamie Boyd, Tommaso Lari and Jalal 
Abdallah for additional information on the ATLAS analyses. We appreciated greatly the help of 
Sabine Kraml for her comments and suggestions on the manuscript and of Bryan
Zaldivar concerning the monojet analysis. GC would also like to 
thank Mathieu Aurousseau for the help with \texttt{ROOT}. The work of GC is supported 
by the Theory-LHC-France initiative of CNRS/IN2P3. The work of DS is supported by the 
French ANR, project DMAstroLHC, ANR-12-BS05-0006, and by the Investissements d'avenir, Labex ENIGMASS.

\clearpage
\appendix

\section{Cutflow for the ATLAS monojet analysis\label{app_monojet}}
We give below the cut flow table for the comparison between the ATLAS monojet analysis 
and the recasted \texttt{MA5} one. The monojets analysis targets the flavour-violating 
decay $\tilde{t} \rightarrow c \tilde{\chi}^{0}_1$ with a branching fraction of 
100\% using a monojet and c-tagged strategies. We only recasted the monojet strategy. 
The dataset corresponds to 20.3 fb$\rm ^{-1}$ of integrated luminosity at $\rm 
\sqrt{s}=8~TeV$.
\begin{table}[!ht]
        \renewcommand\arraystretch{1.2}
        \begin{center}
        \begin{tabular}{|l||c|c||c|c||c|c||}
        \hline
        Benchmark $(\mgino,\mneut{1})$&\multicolumn{2}{c||}{$(200/125)$ GeV} &\multicolumn{2}{c||}{$(200/195)$ GeV}&\multicolumn{2}{c||}{$(250/245)$ GeV}   \\ 
       & {\sc MA}\,5 & CMS & {\sc MA}\,5 & CMS & {\sc MA}\,5 & CMS \\ 
cut    & result & result & result & result & result & result\\ 
\hline
$E_{T}^{\rm miss} > 80$~GeV Filter & $192812.8$ & $181902.0$ & $104577.6$ & $103191.0$& $36055.4$ & $48103.0$   \\ 
$E_{T}^{\rm miss} > 100$~GeV       & $136257.1$ & $97217.0$ & $82619.0$ & $64652.0$ & $29096.3$ & $23416.0$   \\ 
Event cleaning                     & - & $82131.0$ &- & $57566.0$ & - & $21023.0$ \\
Lepton veto & $134894.2$ & $81855.0$ & $82493.9$ & $57455.0$ & $29041.8$ & $20986.0$ \\ 
$N_{\rm jets} \leq 3$ & $101653.7$ & $59315.0$ & $75391.5$ & $52491.0$ & $26295.2$& $18985.0$  \\ 
$\Delta\phi(E_{T}^{\rm miss},{\rm jets}) > 0.4$ & $95568.8$ & $54295.0$ & $70888.1$ & $49216.0$ & $24676.9$  & $17843.0$  \\ 
$p_{T}(j_1) > 150$~GeV & $17282.8$& $14220.0$ & $25552.0$ & $20910.0$ & $9652.1$ & $8183.0$ \\ 
$E_{T}^{\rm miss} > 150$~GeV & $10987.8$ & $9468.0$  & $21569.1$& $18297.0$ & $8363.0$ & $7290.0$ \\
\hline
\multicolumn{7}{|c|}{M1 Signal Region}\\
\hline
$p_{T}(j_1) > 280$~GeV & $2031.2$ & $1627.0$ & $4922.0$ & $3854.0$ & $2156.1$ & $1748.0$ \\ 
$E_{T}^{\rm miss} > 220$~GeV & $1517.6$ & $1276.0$ & $4628.4$ & $3722.0$ & $2022.9$ & $1694.0$  \\ 
\hline
\multicolumn{7}{|c|}{M2 Signal Region}\\
\hline
$p_{T}(j_1) > 340$~GeV & $858.0$  & $721.0$ & $2509.0$& $1897.0$ & $1107.4$& $882.0$\\ 
$E_{T}^{\rm miss} > 340$~GeV & $344.4$ & $282.0$ & $1758.9$ & $1518.0$ & $817.5$& $736.0$ \\ 
\hline
\multicolumn{7}{|c|}{M3 Signal Region}\\
\hline
$p_{T}(j_1) > 450$~GeV & $204.3$ & $169.0$ & $773.3$ & $527.0$ & $376.1$ & $279.0$ \\ 
$E_{T}^{\rm miss} > 450$~GeV & $61.3$ & $64.0$ & $476.8$ & $415.0$ & $268.0$ & $230.0$ \\ 
\hline 
\end{tabular}
\end{center}

\caption{\label{mono_cutflow}\em Cutflow comparison for three benchmark point of the monojet analysis targeting $\tilde{t} \rightarrow c \tilde{\chi}^{0}_1$ in \cite{Aad:2014nra}. The official ATLAS numbers are taken from \cite{mono_twiki}}

\end{table}

\section{Cutflow for the ATLAS multi-jet analysis}
This ATLAS analysis \cite{Aad:2014wea}  searches
for new physics in the 0 lepton + multi-jets (2-6 jets) + missing energy ($ E_{T}^{miss}$) final state.
The dataset corresponds to 20.3 fb$\rm ^{-1}$ of integrated luminosity at $\rm \sqrt{s}=8~TeV$.
 In the context of supersymmetry, the analysis targets  gluino
pair production ($\gino\gino$), squark pair production ($\tilde{q}\tilde{q}$),
 and squark gluino production($\tilde{q}\tilde{g} $). For each of the production modes the following cases are investigated;
\begin{itemize}
\item Gluino pair production :
\begin{enumerate}
\item Direct: $\gino\to q\bar{q}\tilde{\chi}_{1}^{0}$.
\item One step:  $\gino\to q q'\charg{1}$ followed by $\charg{1}\to W \neut{1}$
\end{enumerate}
\item Squark pair production :
\begin{enumerate}
\item Direct : $\tilde{q} \to q \neut{1}$.
\item One step : $\tilde{q}\to  q'\charg{1}$ followed by $\charg{1}\to W \neut{1}$
\end{enumerate}
\item Squark gluino production :
\begin{enumerate}
\item Direct : $\gino\to q\bar{q}\neut{1}$, $\tilde{q} \to q\neut{1}$.
\end{enumerate}
\end{itemize}
\par\noi
For all of the above cases, all branching ratios are assumed to be 100 $\%$
and the rest of the spectrum is decoupled from this set.
\par\noi 
In the following tables we summarize the benchmark points used (Table \ref{tab:spectrumatlas}), the cuts
used for different signal regions(up to 4 jets, Table \ref{tab:cuts}) , and finally
  the comparison between the ATLAS multi-jet analysis and the recasted \texttt{MA5} one (Tables \ref{tab:cutflow1}--\ref{tab:cutflow2}).
Further details about the validation can be found on the web page \cite{MADPAD}.
\begin{table}
\begin{center}
\renewcommand\arraystretch{1.2}
\begin{tabular} {|l|l|l|l |l|l|l|}
\hline
& 2jm  & 2jt & 3j & 4jl- & 4jl & 4jt  \\
decay to LSP &($\tilde{q}\tilde{q}$)  &($\tilde{q}\tilde{q}$)  &($\tilde{g}\tilde{q}$)  &($\tilde{q}\tilde{q}$)  & ($\tilde{g}\tilde{g}$)  & ($\tilde{g}\tilde{g}$)   \\
\hline
&direct & direct & direct & direct & direct & direct  \\
\hline
 $m_{\tilde{q}/\tilde{g}}$ & 475 &  1000  &  1612  &  400    &  800     &  1425      \\
\hline
 $\mcharg{1}$ &--  & --  & -- &--  &--   &-- \\
\hline
$\mneut{1}$ &   425    &  100   &   337   &   250   &   650   &  75     \\ 
\hline
\end{tabular}
\caption{ \it{ The benchmark scenarios used for the valiaion of the 10
signal regions. For the direct squark pair or gluino pair production, the rest of the spectrum is decoupled. 
For squark-gluino production, the gluino and squark masses are assumed to be degenerate.
  All masses are in GeV. The SLHA files were obtained from  \url{http://hepdata.cedar.ac.uk/view/ins1298722}}.
\label{tab:spectrumatlas}}
\end{center}
\end{table}

\begin{table}
\begin{center}
\renewcommand\arraystretch{1.2}
\begin{tabular} {|l|l |l|l|l|l|l|}
\hline
& 2jm  & 2jt & 3j & 4jl- & 4jl & 4jt \\
\hline
 $ E_{T}^{\rm miss}/\sqrt{H_{T}}$ & 15 & 15 &--& 10 & 10 &--    \\
\hline
 $ E_{T}^{\rm miss}/M_{\rm eff}(N_j)$  & -- & -- & 0.3 & --&-- &--  \\
\hline
 $ M_{\rm eff}({\rm incl})$ & 1200 & 1600 & 2200 & 700& 1000 & 2200  \\
\hline
\end{tabular}
\caption{ \it{ The signal region specific cuts. All energy units are in GeV.}
\label{tab:cuts}}
\end{center}
\end{table}

\begin{table}
\begin{center}
\renewcommand\arraystretch{1.2}
\begin{tabular}{| l ||c|c||c|c||c|c|}
\hline
Benchmark/SR & \multicolumn{2}{|c||}{ 2jm ($\tilde{q}\tilde{q}$)} & \multicolumn{2}{c||}{2jt($\tilde{q}\tilde{q}$)} &\multicolumn{2}{c|}{3j($\tilde{g}\tilde{q}$)} \\

    & {\sc MA}\,5  & ATLAS  & {\sc MA}\,5  & ATLAS  & {\sc MA}\,5 & ATLAS   \\ 
cut & result & result       & result & result       & result & result \\
\hline 
$ E_{T}^{\rm miss}>$  160,  &  1656.1 & 1781.2   &62.1  & 61.6  & 18.8& 18.6  \\
 $ p_{T}(j_1,j_2)>$  130,60  &   & & & & &   \\ 
\hline
$ p_{T}(j_3)> 60 $  &--  &--  &--  &-- & 15.1 & 14.8    \\
\hline
$ \Delta\phi(j _{i},E_{T}^{\rm miss}) > 0.4$& 1295.9  & 1462.7& 56.9 & 55.7& 13.3  &12.9   \\
\hline
$ E_{T}^{\rm miss}/\sqrt{H_{T}}   $        & 449.1  & 566.1  & 40.1 & 38.5   & -- & -- \\
\hline
$ E_{T}^{\rm miss}/M_{\rm eff}(N_j) $    & -- & -- & -- & --& 10.1  & 9.6 \\
\hline
$ M_{\rm eff}({\rm incl})$& 122.2   & 102.4& 23.8   & 21.7& 6.2  & 5.9  \\  
\hline\noalign{\smallskip}
\end{tabular}
\caption{ \it{ Cutflows for signal regions 2jm,2jt and 3j, compared to the official ATLAS results
documented in \cite{Aad:2014wea}. All energy units are in GeV.}
\label{tab:cutflow1}}
\end{center}
\end{table}


\begin{table}
\begin{center}
\renewcommand\arraystretch{1.2}
\begin{tabular}{| l ||c|c||c|c||c|c|}
\hline
Benchmark/SR & \multicolumn{2}{|c||}{ 4jl-($\tilde{q}\tilde{q}$)} & \multicolumn{2}{c||}{4jl($\tilde{g}\tilde{g}$)} &\multicolumn{2}{c|}{4jt($\tilde{g}\tilde{g}$)} \\
    & ATLAS  & {\sc MA}\,5  & ATLAS  & {\sc MA}\,5  & ATLAS  & {\sc MA}\,5  \\
cut & result & result       & result & result       & result & result \\
\hline
$ E_{T}^{\rm miss}>$  160, & 16135.8  & 15097   & 634.6   & 679.0   & 13.2   & 12.7 \\
 $ p_{T}(j_1,j_2)>$  130,60  &   & & & &  & \\
\hline
$ p_{T}(j_3,j_4)> 60 $ &  2331 & 2112   & 211.4 & 185.7 & 12.0  & 12.0  \\
\hline
$ \Delta\phi(j _{i\le 3},E_{T}^{\rm miss}) > 0.4$ & 1813.7   & 1723.0   & 154.6  & 144.9  & 8.4   & 8.9   \\
 $ \Delta\phi(j_4,E_{T}^{\rm miss}) > 0.2$  &  & & & & &   \\
\hline
$ E_{T}^{\rm miss}/\sqrt{H_{T}}  $   & 1009     & 943     & 98.7  & 84.4   & -- & -- \\
\hline
$ E_{T}^{\rm miss}/M_{eff}(N_j) $    & -- & -- & -- & -- & 4.8  & 5.5  \\
\hline
$\rm M_{eff}(incl)$ & 884   & 843  & 39.5  & 41.5  & 2.5  &  2.9  \\
\hline\noalign{\smallskip}
\end{tabular}
\caption{ \it{ Cutflows for signal regions 4jl-,4jl and 4jt, compared to the official ATLAS results
documented in \cite{Aad:2014wea}}. 
\label{tab:cutflow2}}
\end{center}
\end{table}


\clearpage

\bibliography{gluino_rad_v2}{}

\providecommand{\href}[2]{#2}\begingroup\raggedright\begin{thebibliography}{100}

\bibitem{atlas-susy-twiki}
{https://twiki.cern.ch/twiki/bin/view/AtlasPublic/SupersymmetryPublicResults}.

\bibitem{cms-susy-twiki}
{https://twiki.cern.ch/twiki/bin/view/CMSPublic/PhysicsResultsSUS}.

\bibitem{atlas-exo-twiki}
{https://twiki.cern.ch/twiki/bin/view/AtlasPublic/ExoticsPublicResults}.

\bibitem{cms-exo-twiki}
{https://twiki.cern.ch/twiki/bin/view/CMSPublic/PhysicsResultsEXO}.

\bibitem{Aad:2015zhl}
{\bfseries ATLAS, CMS} Collaboration, G.~Aad {\em et~al.}, ``{Combined
  Measurement of the Higgs Boson Mass in $pp$ Collisions at $\sqrt{s}=7$ and 8
  TeV with the ATLAS and CMS Experiments},''
  \href{http://dx.doi.org/10.1103/PhysRevLett.114.191803}{{\em Phys. Rev.
  Lett.} {\bfseries 114} (2015) 191803},
\href{http://arxiv.org/abs/1503.07589}{{\ttfamily arXiv:1503.07589 [hep-ex]}}.

\bibitem{Aad:2015iea}
{\bfseries ATLAS} Collaboration, G.~Aad {\em et~al.}, ``{Summary of the
  searches for squarks and gluinos using $\sqrt{s}$ = 8 TeV pp collisions with
  the ATLAS experiment at the LHC},''
\href{http://arxiv.org/abs/1507.05525}{{\ttfamily arXiv:1507.05525 [hep-ex]}}.

\bibitem{deJong:2012zt}
{\bfseries ATLAS, CMS} Collaboration, P.~de~Jong, ``{Supersymmetry searches at
  the LHC},'' in {\em {Proceedings, 32nd International Symposium on Physics in
  Collision (PIC 2012)}}, pp.~241--254.
\newblock 2012.
\newblock \href{http://arxiv.org/abs/1211.3887}{{\ttfamily arXiv:1211.3887
  [hep-ex]}}.
\newblock
\url{https://inspirehep.net/record/1202990/files/arXiv:1211.3887.pdf}.
\newblock

\bibitem{Okawa:2011xg}
{\bfseries ATLAS} Collaboration, H.~Okawa, ``{Interpretations of SUSY Searches
  in ATLAS with Simplified Models},'' in {\em {Particles and fields.
  Proceedings, Meeting of the Division of the American Physical Society, DPF
  2011, Providence, USA, August 9-13, 2011}}.
\newblock 2011.
\newblock \href{http://arxiv.org/abs/1110.0282}{{\ttfamily arXiv:1110.0282
  [hep-ex]}}.
\newblock
\url{https://inspirehep.net/record/930327/files/arXiv:1110.0282.pdf}.
\newblock

\bibitem{Chatrchyan:2013sza}
{\bfseries CMS} Collaboration, S.~Chatrchyan {\em et~al.}, ``{Interpretation of
  Searches for Supersymmetry with simplified Models},''
  \href{http://dx.doi.org/10.1103/PhysRevD.88.052017}{{\em Phys. Rev.}
  {\bfseries D88} no.~5, (2013) 052017},
\href{http://arxiv.org/abs/1301.2175}{{\ttfamily arXiv:1301.2175 [hep-ex]}}.

\bibitem{Alwall:2008ag}
J.~Alwall, P.~Schuster, and N.~Toro, ``{Simplified Models for a First
  Characterization of New Physics at the LHC},''
  \href{http://dx.doi.org/10.1103/PhysRevD.79.075020}{{\em Phys. Rev.}
  {\bfseries D79} (2009) 075020},
\href{http://arxiv.org/abs/0810.3921}{{\ttfamily arXiv:0810.3921 [hep-ph]}}.

\bibitem{Alves:2011sq}
D.~S.~M. Alves, E.~Izaguirre, and J.~G. Wacker, ``{Where the Sidewalk Ends:
  Jets and Missing Energy Search Strategies for the 7 TeV LHC},''
  \href{http://dx.doi.org/10.1007/JHEP10(2011)012}{{\em JHEP} {\bfseries 10}
  (2011) 012},
\href{http://arxiv.org/abs/1102.5338}{{\ttfamily arXiv:1102.5338 [hep-ph]}}.

\bibitem{Alves:2011wf}
{\bfseries LHC New Physics Working Group} Collaboration, D.~Alves,
  ``{Simplified Models for LHC New Physics Searches},''
  \href{http://dx.doi.org/10.1088/0954-3899/39/10/105005}{{\em J. Phys.}
  {\bfseries G39} (2012) 105005},
\href{http://arxiv.org/abs/1105.2838}{{\ttfamily arXiv:1105.2838 [hep-ph]}}.

\bibitem{Ghosh:2012wb}
D.~Ghosh and D.~Sengupta, ``{Searching the sbottom in the four lepton channel
  at the LHC},'' \href{http://dx.doi.org/10.1140/epjc/s10052-013-2342-9}{{\em
  Eur. Phys. J.} {\bfseries C73} no.~3, (2013) 2342},
\href{http://arxiv.org/abs/1209.4310}{{\ttfamily arXiv:1209.4310 [hep-ph]}}.

\bibitem{Dreiner:2012gx}
H.~K. Dreiner, M.~Kramer, and J.~Tattersall, ``{How low can SUSY go? Matching,
  monojets and compressed spectra},''
  \href{http://dx.doi.org/10.1209/0295-5075/99/61001}{{\em Europhys. Lett.}
  {\bfseries 99} (2012) 61001},
\href{http://arxiv.org/abs/1207.1613}{{\ttfamily arXiv:1207.1613 [hep-ph]}}.

\bibitem{Alwall:2008ve}
J.~Alwall, M.-P. Le, M.~Lisanti, and J.~G. Wacker, ``{Searching for Directly
  Decaying Gluinos at the Tevatron},''
  \href{http://dx.doi.org/10.1016/j.physletb.2008.06.065}{{\em Phys. Lett.}
  {\bfseries B666} (2008) 34--37},
\href{http://arxiv.org/abs/0803.0019}{{\ttfamily arXiv:0803.0019 [hep-ph]}}.

\bibitem{Izaguirre:2010nj}
E.~Izaguirre, M.~Manhart, and J.~G. Wacker, ``{Bigger, Better, Faster, More at
  the LHC},'' \href{http://dx.doi.org/10.1007/JHEP12(2010)030}{{\em JHEP}
  {\bfseries 12} (2010) 030},
\href{http://arxiv.org/abs/1003.3886}{{\ttfamily arXiv:1003.3886 [hep-ph]}}.

\bibitem{Alwall:2008va}
J.~Alwall, M.-P. Le, M.~Lisanti, and J.~G. Wacker, ``{Model-Independent Jets
  plus Missing Energy Searches},''
  \href{http://dx.doi.org/10.1103/PhysRevD.79.015005}{{\em Phys. Rev.}
  {\bfseries D79} (2009) 015005},
\href{http://arxiv.org/abs/0809.3264}{{\ttfamily arXiv:0809.3264 [hep-ph]}}.

\bibitem{LeCompte:2011cn}
T.~J. LeCompte and S.~P. Martin, ``{Large Hadron Collider reach for
  supersymmetric models with compressed mass spectra},''
  \href{http://dx.doi.org/10.1103/PhysRevD.84.015004}{{\em Phys. Rev.}
  {\bfseries D84} (2011) 015004},
\href{http://arxiv.org/abs/1105.4304}{{\ttfamily arXiv:1105.4304 [hep-ph]}}.

\bibitem{LeCompte:2011fh}
T.~J. LeCompte and S.~P. Martin, ``{Compressed supersymmetry after 1/fb at the
  Large Hadron Collider},''
  \href{http://dx.doi.org/10.1103/PhysRevD.85.035023}{{\em Phys. Rev.}
  {\bfseries D85} (2012) 035023},
\href{http://arxiv.org/abs/1111.6897}{{\ttfamily arXiv:1111.6897 [hep-ph]}}.

\bibitem{Bhattacherjee:2012mz}
B.~Bhattacherjee and K.~Ghosh, ``{Degenerate SUSY search at the 8 TeV LHC},''
\href{http://arxiv.org/abs/1207.6289}{{\ttfamily arXiv:1207.6289 [hep-ph]}}.

\bibitem{Aad:2014lra}
{\bfseries ATLAS} Collaboration, G.~Aad {\em et~al.}, ``{Search for strong
  production of supersymmetric particles in final states with missing
  transverse momentum and at least three $b$-jets at $\sqrt{s}$= 8 TeV
  proton-proton collisions with the ATLAS detector},''
  \href{http://dx.doi.org/10.1007/JHEP10(2014)024}{{\em JHEP} {\bfseries 10}
  (2014) 24},
\href{http://arxiv.org/abs/1407.0600}{{\ttfamily arXiv:1407.0600 [hep-ex]}}.

\bibitem{Aad:2013wta}
{\bfseries ATLAS} Collaboration, G.~Aad {\em et~al.}, ``{Search for new
  phenomena in final states with large jet multiplicities and missing
  transverse momentum at $\sqrt{s}$=8 TeV proton-proton collisions using the
  ATLAS experiment},'' \href{http://dx.doi.org/10.1007/JHEP10(2013)130,
  10.1007/JHEP01(2014)109}{{\em JHEP} {\bfseries 10} (2013) 130},
  \href{http://arxiv.org/abs/1308.1841}{{\ttfamily arXiv:1308.1841 [hep-ex]}}.
[Erratum: JHEP01,109(2014)].

\bibitem{Chatrchyan:2013iqa}
{\bfseries CMS} Collaboration, S.~Chatrchyan {\em et~al.}, ``{Search for
  supersymmetry in pp collisions at $\sqrt{s}$=8 TeV in events with a single
  lepton, large jet multiplicity, and multiple b jets},''
  \href{http://dx.doi.org/10.1016/j.physletb.2014.04.023}{{\em Phys. Lett.}
  {\bfseries B733} (2014) 328--353},
\href{http://arxiv.org/abs/1311.4937}{{\ttfamily arXiv:1311.4937 [hep-ex]}}.

\bibitem{Chatrchyan:2013fea}
{\bfseries CMS} Collaboration, S.~Chatrchyan {\em et~al.}, ``{Search for new
  physics in events with same-sign dileptons and jets in pp collisions at
  $\sqrt{s}$ = 8 TeV},'' \href{http://dx.doi.org/10.1007/JHEP01(2015)014,
  10.1007/JHEP01(2014)163}{{\em JHEP} {\bfseries 01} (2014) 163},
  \href{http://arxiv.org/abs/1311.6736}{{\ttfamily arXiv:1311.6736}}.
[Erratum: JHEP01,014(2015)].

\bibitem{CMS:2014wsa}
{\bfseries CMS} Collaboration, C.~Collaboration,
``{Exclusion limits on gluino and top-squark pair production in natural SUSY
  scenarios with inclusive razor and exclusive single-lepton searches at 8
  TeV.},''.

\bibitem{Carena:2008mj}
M.~Carena, A.~Freitas, and C.~E.~M. Wagner, ``{Light Stop Searches at the LHC
  in Events with One Hard Photon or Jet and Missing Energy},''
  \href{http://dx.doi.org/10.1088/1126-6708/2008/10/109}{{\em JHEP} {\bfseries
  10} (2008) 109},
\href{http://arxiv.org/abs/0808.2298}{{\ttfamily arXiv:0808.2298 [hep-ph]}}.

\bibitem{Alvarez:2012wf}
E.~Alvarez and Y.~Bai, ``{Reach the Bottom Line of the Sbottom Search},''
  \href{http://dx.doi.org/10.1007/JHEP08(2012)003}{{\em JHEP} {\bfseries 08}
  (2012) 003},
\href{http://arxiv.org/abs/1204.5182}{{\ttfamily arXiv:1204.5182 [hep-ph]}}.

\bibitem{Dreiner:2012sh}
H.~Dreiner, M.~Kraemer, and J.~Tattersall, ``{Exploring QCD uncertainties when
  setting limits on compressed supersymmetric spectra},''
  \href{http://dx.doi.org/10.1103/PhysRevD.87.035006}{{\em Phys. Rev.}
  {\bfseries D87} no.~3, (2013) 035006},
\href{http://arxiv.org/abs/1211.4981}{{\ttfamily arXiv:1211.4981 [hep-ph]}}.

\bibitem{Arbey:2015hca}
A.~Arbey, M.~Battaglia, and F.~Mahmoudi, ``{Monojet Searches for MSSM
  Simplified Models},''
\href{http://arxiv.org/abs/1506.02148}{{\ttfamily arXiv:1506.02148 [hep-ph]}}.

\bibitem{Chatterjee:2014uda}
A.~Chatterjee, A.~Choudhury, A.~Datta, and B.~Mukhopadhyaya, ``{Gluino mass
  limits with sbottom NLSP in coannihilation scenarios},''
  \href{http://dx.doi.org/10.1007/JHEP01(2015)154}{{\em JHEP} {\bfseries 01}
  (2015) 154},
\href{http://arxiv.org/abs/1411.6467}{{\ttfamily arXiv:1411.6467 [hep-ph]}}.

\bibitem{He:2011tp}
B.~He, T.~Li, and Q.~Shafi, ``{Impact of LHC Searches on NLSP Top Squark and
  Gluino Mass},'' \href{http://dx.doi.org/10.1007/JHEP05(2012)148}{{\em JHEP}
  {\bfseries 05} (2012) 148},
\href{http://arxiv.org/abs/1112.4461}{{\ttfamily arXiv:1112.4461 [hep-ph]}}.

\bibitem{Drees:2012dd}
M.~Drees, M.~Hanussek, and J.~S. Kim, ``{Light Stop Searches at the LHC with
  Monojet Events},'' \href{http://dx.doi.org/10.1103/PhysRevD.86.035024}{{\em
  Phys. Rev.} {\bfseries D86} (2012) 035024},
\href{http://arxiv.org/abs/1201.5714}{{\ttfamily arXiv:1201.5714 [hep-ph]}}.

\bibitem{Aad:2015pfx}
{\bfseries ATLAS} Collaboration, G.~Aad {\em et~al.}, ``{ATLAS Run 1 searches
  for direct pair production of third-generation squarks at the Large Hadron
  Collider},''
\href{http://arxiv.org/abs/1506.08616}{{\ttfamily arXiv:1506.08616 [hep-ex]}}.

\bibitem{CMS:2014yma}
{\bfseries CMS} Collaboration, C.~Collaboration,
``{Search for top squarks decaying to a charm quark and a neutralino in events
  with a jet and missing transverse momentum},''.

\bibitem{Ferretti:2015ala}
G.~Ferretti, R.~Franceschini, C.~Petersson, and R.~Torre, ``{Light stop squarks
  and b-tagging},'' in {\em {14th Hellenic School and Workshops on Elementary
  Particle Physics and Gravity Corfu, Attiki, Greece, September 3-21, 2014}}.
\newblock 2015.
\newblock \href{http://arxiv.org/abs/1506.00604}{{\ttfamily arXiv:1506.00604
  [hep-ph]}}.
\newblock
\url{https://inspirehep.net/record/1373918/files/arXiv:1506.00604.pdf}.
\newblock

\bibitem{Toharia:2005gm}
M.~Toharia and J.~D. Wells, ``{Gluino decays with heavier scalar
  superpartners},'' \href{http://dx.doi.org/10.1088/1126-6708/2006/02/015}{{\em
  JHEP} {\bfseries 02} (2006) 015},
\href{http://arxiv.org/abs/hep-ph/0503175}{{\ttfamily arXiv:hep-ph/0503175
  [hep-ph]}}.

\bibitem{Ma:1988ns}
E.~Ma and G.-G. Wong, ``{TWO-BODY RADIATIVE GLUINO DECAYS},''
\href{http://dx.doi.org/10.1142/S0217732388001860}{{\em Mod. Phys. Lett.}
  {\bfseries A3} (1988) 1561}.

\bibitem{Barbieri:1987ed}
R.~Barbieri, G.~Gamberini, G.~F. Giudice, and G.~Ridolfi, ``{Constraining
  Supergravity Models From Gluino Production},''
\href{http://dx.doi.org/10.1016/0550-3213(88)90160-5}{{\em Nucl. Phys.}
  {\bfseries B301} (1988) 15}.

\bibitem{Haber:1983fc}
H.~E. Haber and G.~L. Kane, ``{Gluino Decays and Experimental Signatures},''
\href{http://dx.doi.org/10.1016/0550-3213(84)90570-4}{{\em Nucl. Phys.}
  {\bfseries B232} (1984) 333}.

\bibitem{Baer:1990sc}
H.~Baer, X.~Tata, and J.~Woodside, ``{Phenomenology of Gluino Decays via Loops
  and Top Quark Yukawa Coupling},''
\href{http://dx.doi.org/10.1103/PhysRevD.42.1568}{{\em Phys. Rev.} {\bfseries
  D42} (1990) 1568--1576}.

\bibitem{Bartl:1990ay}
A.~Bartl, W.~Majerotto, B.~Mosslacher, N.~Oshimo, and S.~Stippel, ``{Gluino and
  squark decays into heavy top quarks},''
\href{http://dx.doi.org/10.1103/PhysRevD.43.2214}{{\em Phys. Rev.} {\bfseries
  D43} (1991) 2214--2222}.

\bibitem{Gambino:2005eh}
P.~Gambino, G.~F. Giudice, and P.~Slavich, ``{Gluino decays in split
  supersymmetry},''
  \href{http://dx.doi.org/10.1016/j.nuclphysb.2005.08.011}{{\em Nucl. Phys.}
  {\bfseries B726} (2005) 35--52},
\href{http://arxiv.org/abs/hep-ph/0506214}{{\ttfamily arXiv:hep-ph/0506214
  [hep-ph]}}.

\bibitem{Sato:2012xf}
R.~Sato, S.~Shirai, and K.~Tobioka, ``{Gluino Decay as a Probe of High Scale
  Supersymmetry Breaking},''
  \href{http://dx.doi.org/10.1007/JHEP11(2012)041}{{\em JHEP} {\bfseries 11}
  (2012) 041},
\href{http://arxiv.org/abs/1207.3608}{{\ttfamily arXiv:1207.3608 [hep-ph]}}.

\bibitem{Ghosh:2012dh}
D.~Ghosh, M.~Guchait, S.~Raychaudhuri, and D.~Sengupta, ``{How Constrained is
  the cMSSM?},'' \href{http://dx.doi.org/10.1103/PhysRevD.86.055007}{{\em Phys.
  Rev.} {\bfseries D86} (2012) 055007},
\href{http://arxiv.org/abs/1205.2283}{{\ttfamily arXiv:1205.2283 [hep-ph]}}.

\bibitem{Chatterjee:2012qt}
R.~M. Chatterjee, M.~Guchait, and D.~Sengupta, ``{Probing Supersymmetry using
  Event Shape variables at 8 TeV LHC},''
  \href{http://dx.doi.org/10.1103/PhysRevD.86.075014}{{\em Phys. Rev.}
  {\bfseries D86} (2012) 075014},
\href{http://arxiv.org/abs/1206.5770}{{\ttfamily arXiv:1206.5770 [hep-ph]}}.

\bibitem{Wells:2003tf}
J.~D. Wells, ``{Implications of supersymmetry breaking with a little hierarchy
  between gauginos and scalars},'' in {\em {11th International Conference on
  Supersymmetry and the Unification of Fundamental Interactions (SUSY 2003)
  Tucson, Arizona, June 5-10, 2003}}.
\newblock 2003.
\newblock
\href{http://arxiv.org/abs/hep-ph/0306127}{{\ttfamily arXiv:hep-ph/0306127
  [hep-ph]}}.
\newblock

\bibitem{Wells:2004di}
J.~D. Wells, ``{PeV-scale supersymmetry},''
  \href{http://dx.doi.org/10.1103/PhysRevD.71.015013}{{\em Phys. Rev.}
  {\bfseries D71} (2005) 015013},
\href{http://arxiv.org/abs/hep-ph/0411041}{{\ttfamily arXiv:hep-ph/0411041
  [hep-ph]}}.

\bibitem{ArkaniHamed:2004yi}
N.~Arkani-Hamed, S.~Dimopoulos, G.~F. Giudice, and A.~Romanino, ``{Aspects of
  split supersymmetry},''
  \href{http://dx.doi.org/10.1016/j.nuclphysb.2004.12.026}{{\em Nucl. Phys.}
  {\bfseries B709} (2005) 3--46},
\href{http://arxiv.org/abs/hep-ph/0409232}{{\ttfamily arXiv:hep-ph/0409232
  [hep-ph]}}.

\bibitem{Giudice:2004tc}
G.~F. Giudice and A.~Romanino, ``{Split supersymmetry},''
  \href{http://dx.doi.org/10.1016/j.nuclphysb.2004.11.048}{{\em Nucl. Phys.}
  {\bfseries B699} (2004) 65--89},
  \href{http://arxiv.org/abs/hep-ph/0406088}{{\ttfamily arXiv:hep-ph/0406088
  [hep-ph]}}.
[Erratum: Nucl. Phys.B706,65(2005)].

\bibitem{ArkaniHamed:2004fb}
N.~Arkani-Hamed and S.~Dimopoulos, ``{Supersymmetric unification without low
  energy supersymmetry and signatures for fine-tuning at the LHC},''
  \href{http://dx.doi.org/10.1088/1126-6708/2005/06/073}{{\em JHEP} {\bfseries
  06} (2005) 073},
\href{http://arxiv.org/abs/hep-th/0405159}{{\ttfamily arXiv:hep-th/0405159
  [hep-th]}}.

\bibitem{Arvanitaki:2012ps}
A.~Arvanitaki, N.~Craig, S.~Dimopoulos, and G.~Villadoro, ``{Mini-Split},''
  \href{http://dx.doi.org/10.1007/JHEP02(2013)126}{{\em JHEP} {\bfseries 02}
  (2013) 126},
\href{http://arxiv.org/abs/1210.0555}{{\ttfamily arXiv:1210.0555 [hep-ph]}}.

\bibitem{Harigaya:2013asa}
K.~Harigaya, M.~Ibe, and T.~T. Yanagida, ``{A Closer Look at Gaugino Masses in
  Pure Gravity Mediation Model/Minimal Split SUSY Model},''
  \href{http://dx.doi.org/10.1007/JHEP12(2013)016}{{\em JHEP} {\bfseries 12}
  (2013) 016},
\href{http://arxiv.org/abs/1310.0643}{{\ttfamily arXiv:1310.0643 [hep-ph]}}.

\bibitem{Hall:2011jd}
L.~J. Hall and Y.~Nomura, ``{Spread Supersymmetry},''
  \href{http://dx.doi.org/10.1007/JHEP01(2012)082}{{\em JHEP} {\bfseries 01}
  (2012) 082},
\href{http://arxiv.org/abs/1111.4519}{{\ttfamily arXiv:1111.4519 [hep-ph]}}.

\bibitem{Ibe:2011aa}
M.~Ibe and T.~T. Yanagida, ``{The Lightest Higgs Boson Mass in Pure Gravity
  Mediation Model},''
  \href{http://dx.doi.org/10.1016/j.physletb.2012.02.034}{{\em Phys. Lett.}
  {\bfseries B709} (2012) 374--380},
\href{http://arxiv.org/abs/1112.2462}{{\ttfamily arXiv:1112.2462 [hep-ph]}}.

\bibitem{Ibe:2012hu}
M.~Ibe, S.~Matsumoto, and T.~T. Yanagida, ``{Pure Gravity Mediation with
  m\_{3/2} = 10-100TeV},''
  \href{http://dx.doi.org/10.1103/PhysRevD.85.095011}{{\em Phys. Rev.}
  {\bfseries D85} (2012) 095011},
\href{http://arxiv.org/abs/1202.2253}{{\ttfamily arXiv:1202.2253 [hep-ph]}}.

\bibitem{Evans:2014xpa}
J.~L. Evans and K.~A. Olive, ``{Universality in Pure Gravity Mediation with
  Vector Multiplets},''
  \href{http://dx.doi.org/10.1103/PhysRevD.90.115020}{{\em Phys. Rev.}
  {\bfseries D90} no.~11, (2014) 115020},
\href{http://arxiv.org/abs/1408.5102}{{\ttfamily arXiv:1408.5102 [hep-ph]}}.

\bibitem{Aad:2013gva}
{\bfseries ATLAS} Collaboration, G.~Aad {\em et~al.}, ``{Search for long-lived
  stopped R-hadrons decaying out-of-time with pp collisions using the ATLAS
  detector},'' \href{http://dx.doi.org/10.1103/PhysRevD.88.112003}{{\em Phys.
  Rev.} {\bfseries D88} no.~11, (2013) 112003},
\href{http://arxiv.org/abs/1310.6584}{{\ttfamily arXiv:1310.6584 [hep-ex]}}.

\bibitem{ATLAS:2014fka}
{\bfseries ATLAS} Collaboration, G.~Aad {\em et~al.}, ``{Searches for heavy
  long-lived charged particles with the ATLAS detector in proton-proton
  collisions at $ \sqrt{s}=8 $ TeV},''
  \href{http://dx.doi.org/10.1007/JHEP01(2015)068}{{\em JHEP} {\bfseries 01}
  (2015) 068},
\href{http://arxiv.org/abs/1411.6795}{{\ttfamily arXiv:1411.6795 [hep-ex]}}.

\bibitem{Feldman:2007fq}
D.~Feldman, Z.~Liu, and P.~Nath, ``{Light Higgses at the Tevatron and at the
  LHC and Observable Dark Matter in SUGRA and D Branes},''
  \href{http://dx.doi.org/10.1016/j.physletb.2008.02.063}{{\em Phys. Lett.}
  {\bfseries B662} (2008) 190--198},
\href{http://arxiv.org/abs/0711.4591}{{\ttfamily arXiv:0711.4591 [hep-ph]}}.

\bibitem{Feldman:2008hs}
D.~Feldman, Z.~Liu, and P.~Nath, ``{Sparticles at the LHC},''
  \href{http://dx.doi.org/10.1088/1126-6708/2008/04/054}{{\em JHEP} {\bfseries
  04} (2008) 054},
\href{http://arxiv.org/abs/0802.4085}{{\ttfamily arXiv:0802.4085 [hep-ph]}}.

\bibitem{Feldman:2009zc}
D.~Feldman, Z.~Liu, and P.~Nath, ``{Gluino NLSP, Dark Matter via Gluino
  Coannihilation, and LHC Signatures},''
  \href{http://dx.doi.org/10.1103/PhysRevD.80.015007}{{\em Phys. Rev.}
  {\bfseries D80} (2009) 015007},
\href{http://arxiv.org/abs/0905.1148}{{\ttfamily arXiv:0905.1148 [hep-ph]}}.

\bibitem{Chen:2010kq}
N.~Chen, D.~Feldman, Z.~Liu, P.~Nath, and G.~Peim, ``{Low Mass Gluino within
  the Sparticle Landscape, Implications for Dark Matter, and Early Discovery
  Prospects at LHC-7},''
  \href{http://dx.doi.org/10.1103/PhysRevD.83.035005}{{\em Phys. Rev.}
  {\bfseries D83} (2011) 035005},
\href{http://arxiv.org/abs/1011.1246}{{\ttfamily arXiv:1011.1246 [hep-ph]}}.

\bibitem{Baer:2006dz}
H.~Baer, A.~Mustafayev, E.-K. Park, S.~Profumo, and X.~Tata, ``{Mixed Higgsino
  dark matter from a reduced SU(3) gaugino mass: Consequences for dark matter
  and collider searches},''
  \href{http://dx.doi.org/10.1088/1126-6708/2006/04/041}{{\em JHEP} {\bfseries
  04} (2006) 041},
\href{http://arxiv.org/abs/hep-ph/0603197}{{\ttfamily arXiv:hep-ph/0603197
  [hep-ph]}}.

\bibitem{Baer:2006ff}
H.~Baer, A.~Mustafayev, S.~Profumo, and X.~Tata, ``{Probing SUSY beyond the
  reach of LEP2 at the Fermilab Tevatron: Low |M(3)| dark matter models},''
  \href{http://dx.doi.org/10.1103/PhysRevD.75.035004}{{\em Phys. Rev.}
  {\bfseries D75} (2007) 035004},
\href{http://arxiv.org/abs/hep-ph/0610154}{{\ttfamily arXiv:hep-ph/0610154
  [hep-ph]}}.

\bibitem{Baer:2008ih}
H.~Baer, A.~Mustafayev, E.-K. Park, and X.~Tata, ``{Collider signals and
  neutralino dark matter detection in relic-density-consistent models without
  universality},'' \href{http://dx.doi.org/10.1088/1126-6708/2008/05/058}{{\em
  JHEP} {\bfseries 05} (2008) 058},
\href{http://arxiv.org/abs/0802.3384}{{\ttfamily arXiv:0802.3384 [hep-ph]}}.

\bibitem{Guchait:2011hj}
M.~Guchait, D.~P. Roy, and D.~Sengupta, ``{Probing a Mixed Neutralino Dark
  Matter Model at the 7 TeV LHC},''
  \href{http://dx.doi.org/10.1103/PhysRevD.85.035024}{{\em Phys. Rev.}
  {\bfseries D85} (2012) 035024},
\href{http://arxiv.org/abs/1109.6529}{{\ttfamily arXiv:1109.6529 [hep-ph]}}.

\bibitem{Gogoladze:2009ug}
I.~Gogoladze, R.~Khalid, and Q.~Shafi, ``{Yukawa Unification and Neutralino
  Dark Matter in SU(4)(c) x SU(2)(L) x SU(2)(R)},''
  \href{http://dx.doi.org/10.1103/PhysRevD.79.115004}{{\em Phys. Rev.}
  {\bfseries D79} (2009) 115004},
\href{http://arxiv.org/abs/0903.5204}{{\ttfamily arXiv:0903.5204 [hep-ph]}}.

\bibitem{Gogoladze:2009bn}
I.~Gogoladze, R.~Khalid, and Q.~Shafi, ``{Coannihilation Scenarios and Particle
  Spectroscopy in SU(4)(c) x SU(2)(L) x SU(2)(R)},''
  \href{http://dx.doi.org/10.1103/PhysRevD.80.095016}{{\em Phys. Rev.}
  {\bfseries D80} (2009) 095016},
\href{http://arxiv.org/abs/0908.0731}{{\ttfamily arXiv:0908.0731 [hep-ph]}}.

\bibitem{Ajaib:2010ne}
M.~Adeel~Ajaib, T.~Li, Q.~Shafi, and K.~Wang, ``{NLSP Gluino Search at the
  Tevatron and early LHC},''
  \href{http://dx.doi.org/10.1007/JHEP01(2011)028}{{\em JHEP} {\bfseries 01}
  (2011) 028},
\href{http://arxiv.org/abs/1011.5518}{{\ttfamily arXiv:1011.5518 [hep-ph]}}.

\bibitem{Raza:2014upa}
S.~Raza, Q.~Shafi, and C.~S. Un, ``{NLSP Gluino and NLSP Stop Scenarios from
  b-tau Yukawa Unification},''
\href{http://arxiv.org/abs/1412.7672}{{\ttfamily arXiv:1412.7672 [hep-ph]}}.

\bibitem{Grajek:2013ola}
P.~Grajek, A.~Mariotti, and D.~Redigolo, ``{Phenomenology of General Gauge
  Mediation in light of a 125 GeV Higgs},''
  \href{http://dx.doi.org/10.1007/JHEP07(2013)109}{{\em JHEP} {\bfseries 07}
  (2013) 109},
\href{http://arxiv.org/abs/1303.0870}{{\ttfamily arXiv:1303.0870 [hep-ph]}}.

\bibitem{Dicus:1989gg}
D.~A. Dicus, S.~Nandi, and J.~Woodside, ``{Collider Signals of a Superlight
  Gravitino},''
\href{http://dx.doi.org/10.1103/PhysRevD.41.2347}{{\em Phys. Rev.} {\bfseries
  D41} (1990) 2347}.

\bibitem{Drees:1990vj}
M.~Drees and J.~Woodside, ``{Signals for a superlight gravitino at the LHC},''
  in {\em {ECFA Large Hadron Collider (LHC) Workshop: Physics and
  Instrumentation Aachen, Germany, October 4-9, 1990}}.
\newblock 1990.
\newblock
\url{http://alice.cern.ch/format/showfull?sysnb=0128115}.
\newblock

\bibitem{Dicus:1996ua}
D.~A. Dicus and S.~Nandi, ``{New collider bound on light gravitino mass},''
  \href{http://dx.doi.org/10.1103/PhysRevD.56.4166}{{\em Phys. Rev.} {\bfseries
  D56} (1997) 4166--4169},
\href{http://arxiv.org/abs/hep-ph/9611312}{{\ttfamily arXiv:hep-ph/9611312
  [hep-ph]}}.

\bibitem{Kim:1997iwa}
J.~Kim, J.~L. Lopez, D.~V. Nanopoulos, R.~Rangarajan, and A.~Zichichi, ``{Light
  gravitino production at hadron colliders},''
  \href{http://dx.doi.org/10.1103/PhysRevD.57.373}{{\em Phys. Rev.} {\bfseries
  D57} (1998) 373--382},
\href{http://arxiv.org/abs/hep-ph/9707331}{{\ttfamily arXiv:hep-ph/9707331
  [hep-ph]}}.

\bibitem{Brignole:1998me}
A.~Brignole, F.~Feruglio, M.~L. Mangano, and F.~Zwirner, ``{Signals of a
  superlight gravitino at hadron colliders when the other superparticles are
  heavy},'' \href{http://dx.doi.org/10.1016/S0550-3213(98)00254-5}{{\em Nucl.
  Phys.} {\bfseries B526} (1998) 136--152},
  \href{http://arxiv.org/abs/hep-ph/9801329}{{\ttfamily arXiv:hep-ph/9801329
  [hep-ph]}}.
[Erratum: Nucl. Phys.B582,759(2000)].

\bibitem{Babu:2005ui}
K.~S. Babu, T.~Enkhbat, and B.~Mukhopadhyaya, ``{Split supersymmetry from
  anomalous U(1)},''
  \href{http://dx.doi.org/10.1016/j.nuclphysb.2005.05.006}{{\em Nucl. Phys.}
  {\bfseries B720} (2005) 47--63},
\href{http://arxiv.org/abs/hep-ph/0501079}{{\ttfamily arXiv:hep-ph/0501079
  [hep-ph]}}.

\bibitem{Klasen:2006kb}
M.~Klasen and G.~Pignol, ``{New Results for Light Gravitinos at Hadron
  Colliders: Tevatron Limits and LHC Perspectives},''
  \href{http://dx.doi.org/10.1103/PhysRevD.75.115003}{{\em Phys. Rev.}
  {\bfseries D75} (2007) 115003},
\href{http://arxiv.org/abs/hep-ph/0610160}{{\ttfamily arXiv:hep-ph/0610160
  [hep-ph]}}.

\bibitem{deAquino:2012ru}
P.~de~Aquino, F.~Maltoni, K.~Mawatari, and B.~Oexl, ``{Light Gravitino
  Production in Association with Gluinos at the LHC},''
  \href{http://dx.doi.org/10.1007/JHEP10(2012)008}{{\em JHEP} {\bfseries 10}
  (2012) 008},
\href{http://arxiv.org/abs/1206.7098}{{\ttfamily arXiv:1206.7098 [hep-ph]}}.

\bibitem{Aad:2012pra}
{\bfseries ATLAS} Collaboration, G.~Aad {\em et~al.}, ``{Searches for heavy
  long-lived sleptons and R-Hadrons with the ATLAS detector in $pp$ collisions
  at $\sqrt{s}=7$ TeV},''
  \href{http://dx.doi.org/10.1016/j.physletb.2013.02.015}{{\em Phys. Lett.}
  {\bfseries B720} (2013) 277--308},
\href{http://arxiv.org/abs/1211.1597}{{\ttfamily arXiv:1211.1597 [hep-ex]}}.

\bibitem{Chatrchyan:2013oca}
{\bfseries CMS} Collaboration, S.~Chatrchyan {\em et~al.}, ``{Searches for
  long-lived charged particles in pp collisions at $\sqrt{s}$=7 and 8 TeV},''
  \href{http://dx.doi.org/10.1007/JHEP07(2013)122}{{\em JHEP} {\bfseries 07}
  (2013) 122},
\href{http://arxiv.org/abs/1305.0491}{{\ttfamily arXiv:1305.0491 [hep-ex]}}.

\bibitem{ATLAS:2014qga}
{\bfseries ATLAS} Collaboration, T.~A. collaboration,
``{Limits on metastable gluinos from ATLAS SUSY searches at 8 TeV},''.

\bibitem{Conte:2012fm}
E.~Conte, B.~Fuks, and G.~Serret, ``{MadAnalysis 5, A User-Friendly Framework
  for Collider Phenomenology},''
  \href{http://dx.doi.org/10.1016/j.cpc.2012.09.009}{{\em Comput. Phys.
  Commun.} {\bfseries 184} (2013) 222--256},
\href{http://arxiv.org/abs/1206.1599}{{\ttfamily arXiv:1206.1599 [hep-ph]}}.

\bibitem{Conte:2014zja}
E.~Conte, B.~Dumont, B.~Fuks, and C.~Wymant, ``{Designing and recasting LHC
  analyses with MadAnalysis 5},''
  \href{http://dx.doi.org/10.1140/epjc/s10052-014-3103-0}{{\em Eur. Phys. J.}
  {\bfseries C74} no.~10, (2014) 3103},
\href{http://arxiv.org/abs/1405.3982}{{\ttfamily arXiv:1405.3982 [hep-ph]}}.

\bibitem{Dumont:2014tja}
B.~Dumont, B.~Fuks, S.~Kraml, S.~Bein, G.~Chalons, E.~Conte, S.~Kulkarni,
  D.~Sengupta, and C.~Wymant, ``{Toward a public analysis database for LHC new
  physics searches using MADANALYSIS 5},''
  \href{http://dx.doi.org/10.1140/epjc/s10052-014-3242-3}{{\em Eur. Phys. J.}
  {\bfseries C75} no.~2, (2015) 56},
\href{http://arxiv.org/abs/1407.3278}{{\ttfamily arXiv:1407.3278 [hep-ph]}}.

\bibitem{Djouadi:2006bz}
A.~Djouadi, M.~M. Muhlleitner, and M.~Spira, ``{Decays of supersymmetric
  particles: The Program SUSY-HIT (SUspect-SdecaY-Hdecay-InTerface)},'' {\em
  Acta Phys. Polon.} {\bfseries B38} (2007) 635--644,
\href{http://arxiv.org/abs/hep-ph/0609292}{{\ttfamily arXiv:hep-ph/0609292
  [hep-ph]}}.

\bibitem{Ellis:2015vaa}
J.~Ellis, F.~Luo, and K.~A. Olive, ``{Gluino Coannihilation Revisited},''
\href{http://arxiv.org/abs/1503.07142}{{\ttfamily arXiv:1503.07142 [hep-ph]}}.

\bibitem{Aad:2014nra}
{\bfseries ATLAS} Collaboration, G.~Aad {\em et~al.}, ``{Search for
  pair-produced third-generation squarks decaying via charm quarks or in
  compressed supersymmetric scenarios in $pp$ collisions at $\sqrt{s}=8~$TeV
  with the ATLAS detector},''
  \href{http://dx.doi.org/10.1103/PhysRevD.90.052008}{{\em Phys. Rev.}
  {\bfseries D90} no.~5, (2014) 052008},
\href{http://arxiv.org/abs/1407.0608}{{\ttfamily arXiv:1407.0608 [hep-ex]}}.

\bibitem{Aad:2014wea}
{\bfseries ATLAS} Collaboration, G.~Aad {\em et~al.}, ``{Search for squarks and
  gluinos with the ATLAS detector in final states with jets and missing
  transverse momentum using $\sqrt{s}=8$ TeV proton--proton collision data},''
  \href{http://dx.doi.org/10.1007/JHEP09(2014)176}{{\em JHEP} {\bfseries 09}
  (2014) 176},
\href{http://arxiv.org/abs/1405.7875}{{\ttfamily arXiv:1405.7875 [hep-ex]}}.

\bibitem{Chatrchyan:2014lfa}
{\bfseries CMS} Collaboration, S.~Chatrchyan {\em et~al.}, ``{Search for new
  physics in the multijet and missing transverse momentum final state in
  proton-proton collisions at $\sqrt{s}$= 8 TeV},''
  \href{http://dx.doi.org/10.1007/JHEP06(2014)055}{{\em JHEP} {\bfseries 06}
  (2014) 055},
\href{http://arxiv.org/abs/1402.4770}{{\ttfamily arXiv:1402.4770 [hep-ex]}}.

\bibitem{cms:13012}
S.~Bein and D.~Sengupta,
``{MadAnalysis 5 implementation of CMS-SUS-13-012},''.

\bibitem{atlas:13002}
G.~Chalons and D.~Sengupta,
``{Madanalysis 5 implementation of the ATLAS multi jet analysis documented in
  arXiv:1405.7875, JHEP 1409 (2014) 176},''.

\bibitem{atlas:13021}
D.~Sengupta and G.~Chalons,
``{Madanalysis 5 implementation of the ATLAS monojet analysis documented in
  arXiv:1407.0608, Phys. Rev. D. 90, 052008},''.

\bibitem{mettrigger}
{https://twiki.cern.ch/twiki/bin/view/AtlasPublic/MissingEtTriggerPublicResults}.

\bibitem{MADPAD}
{http://madanalysis.irmp.ucl.ac.be/wiki/PublicAnalysisDatabase}.

\bibitem{Alwall:2011uj}
J.~Alwall, M.~Herquet, F.~Maltoni, O.~Mattelaer, and T.~Stelzer, ``{MadGraph 5
  : Going Beyond},'' \href{http://dx.doi.org/10.1007/JHEP06(2011)128}{{\em
  JHEP} {\bfseries 1106} (2011) 128},
\href{http://arxiv.org/abs/1106.0522}{{\ttfamily arXiv:1106.0522 [hep-ph]}}.

\bibitem{Alwall:2014hca}
J.~Alwall, R.~Frederix, S.~Frixione, V.~Hirschi, F.~Maltoni, O.~Mattelaer,
  H.~S. Shao, T.~Stelzer, P.~Torrielli, and M.~Zaro, ``{The automated
  computation of tree-level and next-to-leading order differential cross
  sections, and their matching to parton shower simulations},''
  \href{http://dx.doi.org/10.1007/JHEP07(2014)079}{{\em JHEP} {\bfseries 07}
  (2014) 079},
\href{http://arxiv.org/abs/1405.0301}{{\ttfamily arXiv:1405.0301 [hep-ph]}}.

\bibitem{Pumplin:2002vw}
J.~Pumplin, D.~R. Stump, J.~Huston, H.~L. Lai, P.~M. Nadolsky, and W.~K. Tung,
  ``{New generation of parton distributions with uncertainties from global QCD
  analysis},'' \href{http://dx.doi.org/10.1088/1126-6708/2002/07/012}{{\em
  JHEP} {\bfseries 07} (2002) 012},
\href{http://arxiv.org/abs/hep-ph/0201195}{{\ttfamily arXiv:hep-ph/0201195
  [hep-ph]}}.

\bibitem{ATL-PHYS-PUB-2011-009}
``{ATLAS tunes of PYTHIA 6 and Pythia 8 for MC11},'' Tech. Rep.
  ATL-PHYS-PUB-2011-009, CERN, Geneva, Jul, 2011.
\newblock \url{https://cds.cern.ch/record/1363300}.

\bibitem{ATL-PHYS-PUB-2011-014}
``{Further ATLAS tunes of PYTHIA6 and Pythia 8},'' Tech. Rep.
  ATL-PHYS-PUB-2011-014, CERN, Geneva, Nov, 2011.
\newblock \url{http://cds.cern.ch/record/1400677}.

\bibitem{Frixione:2002ik}
S.~Frixione and B.~R. Webber, ``{Matching NLO QCD computations and parton
  shower simulations},''
  \href{http://dx.doi.org/10.1088/1126-6708/2002/06/029}{{\em JHEP} {\bfseries
  06} (2002) 029},
\href{http://arxiv.org/abs/hep-ph/0204244}{{\ttfamily arXiv:hep-ph/0204244
  [hep-ph]}}.

\bibitem{Mangano:2006rw}
M.~L. Mangano, M.~Moretti, F.~Piccinini, and M.~Treccani, ``{Matching matrix
  elements and shower evolution for top-quark production in hadronic
  collisions},'' \href{http://dx.doi.org/10.1088/1126-6708/2007/01/013}{{\em
  JHEP} {\bfseries 01} (2007) 013},
\href{http://arxiv.org/abs/hep-ph/0611129}{{\ttfamily arXiv:hep-ph/0611129
  [hep-ph]}}.

\bibitem{Sjostrand:2006za}
T.~Sjostrand, S.~Mrenna, and P.~Z. Skands, ``{PYTHIA 6.4 Physics and Manual},''
  \href{http://dx.doi.org/10.1088/1126-6708/2006/05/026}{{\em JHEP} {\bfseries
  0605} (2006) 026},
\href{http://arxiv.org/abs/hep-ph/0603175}{{\ttfamily arXiv:hep-ph/0603175
  [hep-ph]}}.

\bibitem{deFavereau:2013fsa}
{\bfseries DELPHES 3} Collaboration, J.~de~Favereau, C.~Delaere, P.~Demin,
  A.~Giammanco, V.~Lemaître, A.~Mertens, and M.~Selvaggi, ``{DELPHES 3, A
  modular framework for fast simulation of a generic collider experiment},''
  \href{http://dx.doi.org/10.1007/JHEP02(2014)057}{{\em JHEP} {\bfseries 02}
  (2014) 057},
\href{http://arxiv.org/abs/1307.6346}{{\ttfamily arXiv:1307.6346 [hep-ex]}}.

\bibitem{Kramer:2012bx}
M.~Kramer, A.~Kulesza, R.~van~der Leeuw, M.~Mangano, S.~Padhi, T.~Plehn, and
  X.~Portell, ``{Supersymmetry production cross sections in $pp$ collisions at
  $\sqrt{s}=7$ TeV},''
\href{http://arxiv.org/abs/1206.2892}{{\ttfamily arXiv:1206.2892 [hep-ph]}}.

\bibitem{8tevxs_susy}
{https://twiki.cern.ch/twiki/bin/view/LHCPhysics/ SUSYCrossSections}.

\bibitem{Fox:2011fx}
P.~J. Fox, R.~Harnik, J.~Kopp, and Y.~Tsai, ``{LEP Shines Light on Dark
  Matter},'' \href{http://dx.doi.org/10.1103/PhysRevD.84.014028}{{\em Phys.
  Rev.} {\bfseries D84} (2011) 014028},
\href{http://arxiv.org/abs/1103.0240}{{\ttfamily arXiv:1103.0240 [hep-ph]}}.

\bibitem{Fox:2011pm}
P.~J. Fox, R.~Harnik, J.~Kopp, and Y.~Tsai, ``{Missing Energy Signatures of
  Dark Matter at the LHC},''
  \href{http://dx.doi.org/10.1103/PhysRevD.85.056011}{{\em Phys. Rev.}
  {\bfseries D85} (2012) 056011},
\href{http://arxiv.org/abs/1109.4398}{{\ttfamily arXiv:1109.4398 [hep-ph]}}.

\bibitem{Cacciari:2011ma}
M.~Cacciari, G.~P. Salam, and G.~Soyez, ``{FastJet User Manual},''
  \href{http://dx.doi.org/10.1140/epjc/s10052-012-1896-2}{{\em Eur.Phys.J.}
  {\bfseries C72} (2012) 1896},
\href{http://arxiv.org/abs/1111.6097}{{\ttfamily arXiv:1111.6097 [hep-ph]}}.

\bibitem{Cacciari:2008gp}
M.~Cacciari, G.~P. Salam, and G.~Soyez, ``{The Anti-k(t) jet clustering
  algorithm},'' \href{http://dx.doi.org/10.1088/1126-6708/2008/04/063}{{\em
  JHEP} {\bfseries 0804} (2008) 063},
\href{http://arxiv.org/abs/0802.1189}{{\ttfamily arXiv:0802.1189 [hep-ph]}}.

\bibitem{Borschensky:2014cia}
C.~Borschensky, M.~Kraemer, A.~Kulesza, M.~Mangano, S.~Padhi, {\em et~al.},
  ``{Squark and gluino production cross sections in pp collisions at $\sqrt s=$
  13, 14, 33 and 100 TeV},''
  \href{http://dx.doi.org/10.1140/epjc/s10052-014-3174-y}{{\em Eur.Phys.J.}
  {\bfseries C74} no.~12, (2014) 3174},
\href{http://arxiv.org/abs/1407.5066}{{\ttfamily arXiv:1407.5066 [hep-ph]}}.

\bibitem{Barr:2003rg}
A.~Barr, C.~Lester, and P.~Stephens, ``{m(T2): The Truth behind the glamour},''
  \href{http://dx.doi.org/10.1088/0954-3899/29/10/304}{{\em J.Phys.} {\bfseries
  G29} (2003) 2343--2363},
\href{http://arxiv.org/abs/hep-ph/0304226}{{\ttfamily arXiv:hep-ph/0304226
  [hep-ph]}}.

\bibitem{Barr:2009wu}
A.~J. Barr and C.~Gwenlan, ``{The Race for supersymmetry: Using m(T2) for
  discovery},'' \href{http://dx.doi.org/10.1103/PhysRevD.80.074007}{{\em
  Phys.Rev.} {\bfseries D80} (2009) 074007},
\href{http://arxiv.org/abs/0907.2713}{{\ttfamily arXiv:0907.2713 [hep-ph]}}.

\bibitem{Lester:1999tx}
C.~Lester and D.~Summers, ``{Measuring masses of semiinvisibly decaying
  particles pair produced at hadron colliders},''
  \href{http://dx.doi.org/10.1016/S0370-2693(99)00945-4}{{\em Phys.Lett.}
  {\bfseries B463} (1999) 99--103},
\href{http://arxiv.org/abs/hep-ph/9906349}{{\ttfamily arXiv:hep-ph/9906349
  [hep-ph]}}.

\bibitem{Belanger:2013oka}
G.~Belanger, D.~Ghosh, R.~Godbole, M.~Guchait, and D.~Sengupta, ``{Probing the
  flavor violating scalar top quark signal at the LHC},''
  \href{http://dx.doi.org/10.1103/PhysRevD.89.015003}{{\em Phys. Rev.}
  {\bfseries D89} (2014) 015003},
\href{http://arxiv.org/abs/1308.6484}{{\ttfamily arXiv:1308.6484 [hep-ph]}}.

\bibitem{mono_twiki}
{https://atlas.web.cern.ch/Atlas/GROUPS/PHYSICS/PAPERS/SUSY-2013-21/}.

\end{thebibliography}\endgroup
\bibliographystyle{utphys}

\end{document}